\documentclass[12pt]{article}
\pdfoutput=1
\usepackage[colorlinks,linkcolor=Blue,citecolor=Blue,urlcolor=black,bookmarks,bookmarksnumbered]{hyperref}
\usepackage[scaled=0.85]{helvet}
\usepackage{amsmath,amssymb,accents,mathrsfs,XoohmE}
\usepackage{graphicx,color}
\usepackage{booktabs}
\usepackage{multirow}
\usepackage{placeins}
\usepackage{amsmath}
\usepackage{tensind}
\tensordelimiter{?}

\usepackage{XoohmE}

\definecolor{Green}  {rgb}{0.10,0.70,0.10} 
\definecolor{Orange} {rgb}{1.00,0.50,0.15} 
\definecolor{Red}    {rgb}{0.90,0.00,0.12} 
\definecolor{Purple} {rgb}{0.50,0.25,0.55} 
\definecolor{Turque} {rgb}{0.00,0.65,0.85} 
\definecolor{Blue}   {rgb}{0.00,0.00,1.00} 
\definecolor{Magenta}{rgb}{1.00,0.00,1.00} 
\definecolor{Gold}   {rgb}{1.00,0.75,0.25} 
\definecolor{Seaweed}{rgb}{0.01,0.24,0.09} 
\definecolor{Brown}  {rgb}{0.43,0.26,0.32} 
\definecolor{grey1}  {rgb}{0.20,0.20,0.20} 
\definecolor{grey2}  {rgb}{0.40,0.40,0.40} 
\definecolor{grey3}  {rgb}{0.60,0.60,0.60} 
\definecolor{grey4}  {rgb}{0.80,0.80,0.80} 
\definecolor{grey5}  {rgb}{0.90,0.90,0.90} 
\def\C#1#2{{\ifcase#1\or
             \color{Green}\or \color{Orange}\or \color{Red}\or
              \color{Purple}\or \color{Turque}\or \color{Blue}\or
               \color{Magenta}\or \color{Gold}\or \color{Seaweed}\or
                \color{Brown}\or\color{grey1}\or\color{grey2}\or
                 \color{grey3}\else\color{grey4}\fi#2}}

\definecolor{Slate} {rgb}{0.00,0.45,0.55}

\def\fracm#1#2{\hbox{\large{${\frac{{#1}}{{#2}}}$}}}

\def\be{\begin{equation}}
\def\ee{\end{equation}}
\newcommand{\bea}{\begin{eqnarray}}
\newcommand{\eea}{\end{eqnarray}}
\newcommand{\ena}{\end{eqnarray}}

\def\pp{{\mathchoice
          {
              \kern 1pt%
              \raise 1pt
              \vbox{\hrule width5pt height0.4pt depth0pt
                    \kern -2pt
                    \hbox{\kern 2.3pt
                          \vrule width0.4pt height6pt depth0pt
                          }
                    \kern -2pt
                    \hrule width5pt height0.4pt depth0pt}%
                    \kern 1pt
           }
            {
              \kern 1pt%
              \raise 1pt
              \vbox{\hrule width4.3pt height0.4pt depth0pt
                    \kern -1.8pt
                    \hbox{\kern 1.95pt
                          \vrule width0.4pt height5.4pt depth0pt
                          }
                    \kern -1.8pt
                    \hrule width4.3pt height0.4pt depth0pt}%
                    \kern 1pt
            }
            {
              \kern 0.5pt%
              \raise 1pt
              \vbox{\hrule width4.0pt height0.3pt depth0pt
                    \kern -1.9pt  
                    \hbox{\kern 1.85pt
                          \vrule width0.3pt height5.7pt depth0pt
                          }
                    \kern -1.9pt
                    \hrule width4.0pt height0.3pt depth0pt}%
                    \kern 0.5pt
            }
            {
              \kern 0.5pt%
              \raise 1pt
              \vbox{\hrule width3.6pt height0.3pt depth0pt
                    \kern -1.5pt
                    \hbox{\kern 1.65pt
                          \vrule width0.3pt height4.5pt depth0pt
                          }
                    \kern -1.5pt
                    \hrule width3.6pt height0.3pt depth0pt}%
                    \kern 0.5pt
            }
        }}

\def\mm{{\mathchoice
                       {
                             \kern 1pt
               \raise 1pt    \vbox{\hrule width5pt height0.4pt depth0pt
                                  \kern 2pt
                                  \hrule width5pt height0.4pt depth0pt}
                             \kern 1pt}
                       {
                            \kern 1pt
               \raise 1pt \vbox{\hrule width4.3pt height0.4pt depth0pt
                                  \kern 1.8pt
                                  \hrule width4.3pt height0.4pt depth0pt}
                             \kern 1pt}
                       {
                            \kern 0.5pt
               \raise 1pt
                            \vbox{\hrule width4.0pt height0.3pt depth0pt
                                  \kern 1.9pt
                                  \hrule width4.0pt height0.3pt depth0pt}
                            \kern 1pt}
                       {
                           \kern 0.5pt
             \raise 1pt  \vbox{\hrule width3.6pt height0.3pt depth0pt
                                  \kern 1.5pt
                                  \hrule width3.6pt height0.3pt depth0pt}
                           \kern 0.5pt}
                       }}

\def\ad{{\kern0.5pt
                   \alpha \kern-5.05pt \raise5.8pt\hbox{$\textstyle.$}\kern
0.5pt}}

\def\bd{{\kern0.5pt
                   \beta \kern-5.05pt \raise5.8pt\hbox{$\textstyle.$}\kern
0.5pt}}

\def\qd{{\kern0.5pt
                   q \kern-5.05pt \raise5.8pt\hbox{$\textstyle.$}\kern
0.5pt}}
\def\Dot#1{{\kern0.5pt
     {#1} \kern-5.05pt \raise5.8pt\hbox{$\textstyle.$}\kern
0.5pt}}

\catcode`@=11
\def\un#1{\relax\ifmmode\@@underline#1\else
        $\@@underline{\hbox{#1}}$\relax\fi}
\catcode`@=12

\def\a{\alpha}
\def\b{\beta}
\def\c{\chi}
\def\d{\delta}

\def\g{\gamma}

\def\i{\iota}
\def\j{\psi}
\def\k{\kappa}
\def\l{\lambda}
\def\m{\mu}
\def\n{\nu}
\def\o{\omega}

\def\r{\rho}
\def\s{\sigma}
\def\t{\tau}

\def\L{\Lambda}
\def\O{\Omega}
\def\P{\Pi}

\def\S{\Sigma}

\def\ca{{\cal A}}

\def\cf{{\cal F}}

\def\dslash{\not{\hbox{\kern-2pt $\partial$}}}
\def\Dslash{\not{\hbox{\kern-4pt $D$}}}
\def\pslash{\not{\hbox{\kern-2.3pt $p$}}}
 \newtoks\slashfraction
 \slashfraction={.13}
 \def\slash#1{\setbox0\hbox{$ #1 $}
 \setbox0\hbox to \the\slashfraction\wd0{\hss \box0}/\box0 }
 
\def\kcr{{\hbox{\ro \char'170}}}                
\def\ktl{{\hbox{\ro \char'170}}}        
\def\ktr{{\hbox{\ro \char'170}}}        
\def\kbl{{\hbox{\ro \char'170}}}        
\def\kbr{{\hbox{\ro \char'170}}}        

\def\plpl{\raise-2pt\hbox{$\raise3pt\hbox{$_+$}\hskip-6.67pt\raise0.0pt
\hbox{$^+$}\hskip 0.01pt$}}
\def\mimi{\raise-2pt\hbox{$\raise3pt\hbox{$_-$}\hskip-6.67pt\raise0.0pt
\hbox{$^-$}\hskip 0.01pt$}} 

\def\bo{{\raise.15ex\hbox{\large$\Box$}}}               
\def\pa{\partial}                                       
\def\TH{{\raise.2ex\hbox{$\displaystyle \bigodot$}\mskip-4.7mu \llap H \;}}
\def\face{{\raise.2ex\hbox{$\displaystyle \bigodot$}\mskip-2.2mu \llap {$\ddot
        \smile$}}}                                      

\def\dt#1{\on{\hbox{\bf .}}{#1}}                
\def\Dot#1{\dt{#1}}

\def\Tilde#1{\widetilde{#1}}                    
\def\Hat#1{\widehat{#1}}                        
\def\leftrightarrowfill{$\mathsurround=0pt \mathord\leftarrow \mkern-6mu
        \cleaders\hbox{$\mkern-2mu \mathord- \mkern-2mu$}\hfill
        \mkern-6mu \mathord\rightarrow$}
\def\dvec#1{\vbox{\ialign{##\crcr
        \leftrightarrowfill\crcr\noalign{\kern-1pt\nointerlineskip}
        $\hfil\displaystyle{#1}\hfil$\crcr}}}           
\def\dt#1{{\buildrel {\hbox{\LARGE .}} \over {#1}}}     

\def\fracm#1#2{\hbox{\large{${\frac{{#1}}{{#2}}}$}}}
\def\sfrac#1#2{{\vphantom1\smash{\lower.5ex\hbox{\small$#1$}}\over
        \vphantom1\smash{\raise.4ex\hbox{\small$#2$}}}} 
\def\bfrac#1#2{{\vphantom1\smash{\lower.5ex\hbox{$#1$}}\over
        \vphantom1\smash{\raise.3ex\hbox{$#2$}}}}       
\def\afrac#1#2{{\vphantom1\smash{\lower.5ex\hbox{$#1$}}\over#2}}    

\def\pa{\partial}      
\let\bm\relax
\newcommand{\bm}[1]{{\boldsymbol{#1}}}

\def\ad{{\dot{\alpha}}}
\def\bd{{\dot{\beta}}}

 \font\rOpe=cmsy10                        
 \def\ktl{{\hbox{\rOpe\char'170}}}        
 \def\kbl{{\hbox{\rOpe\char'170}}}        
 \def\kcr{{\reflectbox{\rOpe\char'170}}}        
 \def\ktr{{\reflectbox{\rOpe\char'170}}}        
 \def\kbr{{\reflectbox{\rOpe\char'170}}}        
 \def\Border{\vbox{\hsize0pt
        \setlength{\unitlength}{1mm}
        \newcount\xco
        \newcount\yco
        \xco=-21
        \yco=12
        \begin{picture}(0,0)(-7.5,0)
        \put(\xco,\yco){$\ktl$}
        \advance\yco by-1
        {\loop
        \put(\xco,\yco){$\kcr$}
        \advance\yco by-2
        \ifnum\yco>-240
        \repeat
        \put(\xco,\yco){$\kbl$}}
        \xco=170
        \yco=12
        \put(\xco,\yco){$\ktr$}
        \advance\yco by-1
        {\loop
        \put(\xco,\yco){$\kcr$}
        \advance\yco by-2
        \ifnum\yco>-240
        \repeat
        \put(\xco,\yco){$\kbr$}}
        \put(-19.5,13){\scalebox{.6065}{%
         University of Maryland Center for String and Particle  Theory \&\ Physics Department%
        |University of Maryland Center for String and Particle  Theory \&\ Physics Department}}
        \put(-19.5,-241.5){\scalebox{.5835}{%
         ****University of Maryland * Center for String and
         Particle  Theory* Physics Department****University of Maryland *Center
        for String and Particle  Theory* Physics Department}}
        \end{picture}
        \par\vskip-8mm}}
\definecolor{UMred}{rgb}{.9,.05,.2}
\definecolor{HUblue}{rgb}{.0,.3,.7}

\definecolor{Red}    {rgb}{0.90,0.00,0.12} 
\definecolor{Blue}   {rgb}{0.00,0.00,1.00} 
\definecolor{Green}  {rgb}{0.10,0.70,0.10} 
\definecolor{Turque} {rgb}{0.00,0.65,0.85} 
\definecolor{Orange} {rgb}{1.00,0.50,0.15} 
\definecolor{Magenta}{rgb}{1.00,0.00,1.00} 
\definecolor{Gold}   {rgb}{1.00,0.75,0.25} 
\definecolor{Seaweed}{rgb}{0.01,0.24,0.09} 
\definecolor{Purple} {rgb}{0.50,0.25,0.55} 
\definecolor{Brown}  {rgb}{0.43,0.26,0.32} 
\definecolor{grey1}  {rgb}{0.20,0.20,0.20} 
\definecolor{grey2}  {rgb}{0.40,0.40,0.40} 
\definecolor{grey3}  {rgb}{0.60,0.60,0.60} 
\definecolor{grey4}  {rgb}{0.80,0.80,0.80} 
\definecolor{grey5}  {rgb}{0.90,0.90,0.90} 
\def\C#1#2{{\ifcase#1\or
             \color{Red}\or \color{Green}\or \color{Blue}\or\
              \color{Turque}\or \color{Orange}\or \color{Magenta}\or 
               \color{Gold}\or \color{Seaweed}\or \color{Purple}\or
                \color{Brown}\or\color{grey1}\or\color{grey2}\or
                 \color{grey3}\else\color{grey4}\fi#2}}

\definecolor{Slate} {rgb}{0.00,0.45,0.55}

\newdimen\parshift\parshift=\parindent
\catcode`@=11
 \long\def\@footnotetext#1{\insert\footins{\reset@font\footnotesize
           \interlinepenalty\interfootnotelinepenalty\splittopskip%
            \footnotesep\splitmaxdepth\dp\strutbox\floatingpenalty\@MM%
             \hsize\columnwidth\addtolength{\hsize}{-2\parindent}
              \@parboxrestore\protected@edef\@currentlabel%
              {\csname p@footnote\endcsname\@thefnmark}%
                \color@begingroup%
                 \@makefntext{\rule\z@\footnotesep\ignorespaces#1%
                  \@finalstrut\strutbox}%
                \color@endgroup}}
 \long\def\@makefntext#1{\hglue\parshift%
           \vbox{\noindent\baselineskip=11pt plus.5pt minus.5pt\hb@xt@0em{\hss\@makefnmark\kern1pt}#1}}
\catcode`@=12

\newskip\humongous \humongous=0pt plus 1000pt minus 1000pt
\def\caja{\mathsurround=0pt}
\def\eqalign#1{\,\vcenter{\openup2\jot \caja
        \ialign{\strut \hfil$\displaystyle{##}$&$
        \displaystyle{{}##}$\hfil\crcr#1\crcr}}\,}
\newif\ifdtup

\makeatletter
\def\section{\@startsection{section}{1}{\z@}
        {3ex plus-1ex minus-.2ex}{1pt plus1pt}{\large\sf\bfseries\boldmath}}
\def\subsection{\@startsection{subsection}{2}{\z@}
         {1.5ex plus-1ex minus-.2ex}{0.01pt plus1pt}{\sf\slshape}}
\def\subsubsection{\@startsection{subsubsection}{3}{\z@}
          {1.5ex plus-1ex minus-.2ex}{0.01pt plus0.2pt}{\sf\boldmath}}
\def\paragraph{\@startsection{paragraph}{4}{\z@}
           {.75ex \@plus.5ex \@minus.2ex}{-2mm}{\sf\bfseries\boldmath}}
\makeatother

 \allowdisplaybreaks
 \seceq

\begin{document}

\thispagestyle{empty}
%
\noindent{\small
\hfill{Brown-HET-1780  \\ 
$~~~~~~~~~~~~~~~~~~~~~~~~~~~~~~~~~~~~~~~~~~~~~~~~~~~~~~~~~~~~~~~~~$
$~~~~~~~~~~~~~~~~~~~~~~~~~~~~~~~~~~~~~~~~~~~~~~~~~~~~~~~~~~~~~~~~~$
{}
}
\vspace*{8mm}
\begin{center}
{\large \bf
Exploring the Abelian 4D, $\bm {\cal N}$ = 4 Vector-Tensor
Supermultiplet
\\[6pt] 
and
\\[8pt] 
Its Off-Shell Central Charge Structure}   \\   [12mm]
{\large {
S.\ James Gates, Jr.\footnote{\href{mailto:sylvester\_gates@brown.edu}{sylvester\_gates@brown.edu}}
and Kory Stiffler\footnote{\href{mailto:kory\_stiffler@brown.edu}{kory\_stiffler@brown.edu}}
}}
\\*[12mm]
\emph{
\centering
Department of Physics, Brown University,
\\[1pt]
Box 1843, 182 Hope Street, Barus \& Holley,
Providence, RI 02912, USA 
}
 \\*[20mm]
{ ABSTRACT}\\[4mm]
\parbox{132mm}{\parindent=2pc\indent\baselineskip=14pt plus1pt
An abelian 4D, $\mathcal{N}$ = 4 vector supermultiplet allows for a duality transformation
to be applied to one of its spin-0 states. The resulting theory can be described as an 
abelian 4D, $\mathcal{N}$ = 4 vector-tensor supermultiplet.  It is seen to decompose into 
a direct sum of an off-shell 4D, $\mathcal{N}$ = 2 vector supermultiplet and an off-shell 
4D, $\mathcal{N}$ = 2 tensor supermultiplet. The commutator algebra of the other two 
supersymmetries are still found to be on-shell. However, the central charge structure 
in the resulting 4D, $\mathcal{N}$ = 4 vector-tensor supermultiplet is considerably simpler 
that that of the parent abelian 4D, $\mathcal{N}$ = 4 vector supermultiplet.  This appears to 
be due to the replacement of the usual SO(4) symmetry associated with the abelian 
4D, $\mathcal{N}$ = 4 vector supermultiplet being replaced by a GL(2,$\mathbb{R}$)$\otimes$GL(2,$\mathbb{R}$)
symmetry in the 4D, $\mathcal{N}$ = 4 vector-tensor supermultiplet. The  \emph{Mathematica} code detailing the calculations is available open-source at the \href{https://hepthools.github.io/Data/}{HEPTHools Data Repository} on GitHub.}
 \end{center}
\vfill
\vfill
\clearpage

\section{Introduction}

There currently exists in the physics literature very few examples of four dimensional
relativistic quantum field theories that realize 4D, $\cal N$ = 4 supersymmetry.  In fact,
to our knowledge the only known examples currently in the literature are: (a.) 4D, $\cal N$ 
= 4 supergravity theories \cite{4dn4SG1,4dn4SG2,4dn4SG3}, and (b.) 4D, $\cal N$ = 
4 super Yang-Mills theories \cite{4dn4YMa,4dn4YMb}.  With such a paucity of these 
types of supermultiplets, we believe it might prove useful to cast
the dual version of the 4D, $\cal N$ = 4 super Yang-Mills theories in a new light as pertains to the internal isospin structure. Specifically, in this paper we will focus on the 4D, $\cal N$=4 vector-tensor multiplet originally described in~\cite{Sohnius:1980it}. As is well-known 
the spectrum of 4D, $\cal N$ = 4 abelian super vector supermultiplets contains one 
spin-1 boson and six spin-0 bosons, one can perform a duality transformation on 
one of the spin-0 bosons to replace it by a second rank anti-symmetric tensor.  In~\cite{Sohnius:1980it}, a 4D, $\cal N$ = 4 vector-tensor multiplet was presented with an SP(4) internal symmetry of the scalars and fermions. Effectively, this multiplet was an SP(4) extension of the 4D, $\cal N$=2 vector-tensor multiplet, which has central charges. Dimensionally reducing to the 4D, $\cal N$ = 4 vector-tensor multiplet was discussed in~\cite{Siegel:1980gd}. More recently, other facets of vector-tensor multiplets have been developed such as couplings to supergravity, Chern-Simons, and self-interactions~\cite{Claus:1995sv,Claus:1996ze,Claus:1997fk,Dragon:1997za,Dragon:1997hw,Claus:1998jt}.

In this note, we express the $4D$, $\cal N$ = 4 vector-tensor multiplet in an explicit GL(2,$\mathbb{R}$)$\otimes$GL(2,$\mathbb{R}$) isospin form, where one GL(2,$\mathbb{R}$) corresponds to  a  4D, $\cal N$ = 2 vector supermultiplet and the other GL(2,$\mathbb{R}$) corresponds to a 4D, $\cal{N}$ =2 tensor supermultiplet, neither of which have central charges. The 4D, $\cal N$=2 vector supermultiplet contains a 4D, $\cal N$ = 1 vector 
supermultiplet and a 4D, $\cal N$ = 1 chiral supermultiplet.  In a similar manner, the 
4D, $\cal N$ = 2 tensor supermultiplet contains a 4D, $\cal N$ = 1 tensor supermultiplet 
and a 4D, $\cal N$ = 1 chiral supermultiplet. 

Our examination begins with the 4D, $\cal N$ = 2 vector supermultiplet  and the 4D, $\cal N$ = 2  tensor supermultiplet, with their manifest off-shell 4D, $\cal N$ = 2 supersymmetry, and
constructs from them a 4D, $\cal N$ = 4 vector-tensor supermultiplet. We will examine the resulting supermultiplet in a Majorana notation as in~\cite{Gates:2015dda}, but cast into a GL(2,$\mathbb{R}$)$\otimes$GL(2,$\mathbb{R}$) isospin form where the underlying 4D, $\cal N$=2 supersymmetric tensor and vector supermultiplets take a similar form to those investigated in~\cite{Gates:2014vxa}. 

This paper is organized as follows. The main results of the 4D, $\cal N$ = 4 vector-tensor supermultiplet with GL(2,$\mathbb{R}$)$\otimes$GL(2,$\mathbb{R}$) isospin structure are given in section~\ref{s:vt} and the appendices referenced therein. In contrast, section~\ref{s:cv} and the appendices referenced therein display the usual 4D, $\cal N$ = 4 vector multiplet, expressed in terms of a GL(2,$\mathbb{R}$)$\otimes$GL(2,$\mathbb{R}$) isospin structure. The version of the 4D, $\cal N$ = 4 vector multiplet investigated in~\cite{Gates:2014vxa} is reviewed in appendix~\ref{a:N4Review}. A Majorana notation is used throughout the paper.

\section{Moving from 4D, \texorpdfstring{$\cal N$}{N} = 2 to 4D, \texorpdfstring{$\cal N$}{N} = 4 SUSY }\label{s:vt}

The supersymmetry transformation properties and Lagrangians for both the 4D, $\cal 
N$ = 2 vector super-\\multiplet $\{A,B,F,G, A{}_{\mu}, d,\Psi^{k}_{c}\}$ and the 4D, $\cal 
N$ = 2 tensor supermultiplet $\{{\Tilde A},{\Tilde B},{\Tilde F},{\Tilde G}, {\Tilde {
\varphi}}, B{}_{\mu \nu}, {\Tilde {\Psi}}^{k}_{c}\} $ are given in the appendix.  The 
expectation is that if we add their two respective Lagrangians $\mathcal{L}_{(2VS)}$
and $\mathcal{L}_{(2TS)}$ together, the sum of these should be able to realize two 
additional supersymmetries.  Therefore, we introduce a ``second doublet'' of supersymmetrical
covariant derivative operators denoted by ${\Tilde {\text D}}{}^{i}_{a}$ where we 
use the ``tilde'' in the notation ${\Tilde {\text D}}{}^{i}_{a}$ to distinguish these from
the covariant derivatives ${ {\text D}}{}^{i}_{a}$ associated with the two manifest 
supersymmetries.  The first set of transformations involving ${\text D}_a^i$ is given in appendix~\ref{a:N2Review}. The second supersymmetry covariant derivative ${\Tilde {\text D}}{}^{i}_{a}$  is represented 
by a transformation with the property that under its action, any field in the 4D, $\cal 
N$ = 2 vector supermultiplet is transformed into a field in the 4D, $\cal N$ = 2 tensor 
supermultiplet and vice versa. We also require these transformations to act linearly 
on the field variables.  We are thus motivated to make an ansatz that requires the 
introduction of two sets of twelve matrices in GL(2,$\mathbb{R}$) $\{ ({\cal V}{}_{1}){}^{i j}$, $\dots$, 
$({\cal V}{}_{12}){}^{i j} \}$ and $\{ ({\cal U}{}_{1}){}^{i j}$, $\dots$, $({\cal U}{}_{12}){
}^{i j} \}$ that are used to write a realization of the action of ${\Tilde {\text D}}{}^{i}_{a}$ 
according to 
\vskip0.1mm 
$$
\eqalign{
{\Tilde {\text D}}{}^{i}_{a} A 
~&=~   ({\cal V}{}_{1}){}^{i j} {\Tilde {\Psi}}^{j}_{a}  ~~~, \cr
{\Tilde {\text D}}{}^{i}_{a} B 
~&=~ i  ({\cal V}{}_{2}){}^{i j}(\g^{5})_{a}{}^{b} {\Tilde {\Psi}}^{j}_{b}  ~~~, \cr
{\Tilde {\text D}}{}^{i}_{a} F  ~&=~  ({\cal V}{}_{3}){}^{i j} (\g^{\mu})_{a}{
}^{b} \pa_{\mu} {\Tilde {\Psi}}^{j}_{b}  ~~~,  \cr
{\Tilde {\text D}}{}^{i}_{a} G  ~&=~ i  ({\cal V}{}_{4}){}^{i j} (\g^{5}\g^{
\mu})_{a}{}^{b} \pa_{\mu} {\Tilde {\Psi}}^{j}_{b}  ~~~, \cr
{\Tilde {\text D}}{}^{i}_{a} A_{\mu} ~&=~ i ({\cal V}{}_{5}){}^{ij} (\g_{\mu
})_{a}{}^{b} {\Tilde \Psi}{}^{j}_{b}  ~~~, \cr
 {\Tilde {\text D}}{}^{i}_{a} d  ~&=~ i ({\cal V}{}_{6}){}^{i j} (\g^{5}\g^{
 \mu})_{a}{}^{b} \pa_{\mu} {\Tilde {\Psi}} {}^{j}_{b}  ~~~, 
 {~~~~~~~~~~~~~~~~~~~~~~~~~~~~~~~~~~~~~~}
 } $$
 \be
\begin{split}
 {\Tilde {\text D}}{}^{i}_{a} \Psi^{j}_{b} ~&=~ \big[ -  ({\cal V}{}_{7}){}^{i j}  (\g^{
5}\g^{\mu})_{ab} \pa_{\mu} {\Tilde B}
 - i  ({\cal V}{}_{8}){}^{i j} C_{ab} {\Tilde F} ~~~, \\
 & ~~~~~+  ({\cal V}{}_{9}){}^{i j} (\g^{5})_{ab} {\Tilde G}  \big] + i ({\cal V}{}_{
 10}){}^{i j} (\g^{\mu})_{ab} \pa_{\mu} {\Tilde A}  ~~~, \\
 & ~~~+ i ({\cal V}{}_{11}){}^{i j} (\g^{\mu})_{ab} \pa_{\mu} {\Tilde {\varphi}
 } - i ({\cal V}{}_{12}){}^{i j} \epsilon_{\mu}{}^{\nu\alpha\beta} (\g^{5}\g^{
 \mu})_{ab} \pa_{\nu} {\Tilde B}_{\alpha\beta}  ~~~,
\end{split}
\ee
on the fields in the 4D, $\cal N$ = 2 vector supermultiplet and as
\be
\begin{split}
{\Tilde {\text D}}{}^{i}_{a}{\Tilde A}  ~&=~  ({\cal U}{}_{1}){}^{i j} \Psi^{j}_{a}  ~~~, \\
{\Tilde {\text D}}{}^{i}_{a}{\Tilde B}
~&=~ i ({\cal U}{}_{2}){}^{i j} (\g^{5})_{a}{}^{b} \Psi^{j}_{b}  ~~~, \\
 {\Tilde {\text D}}{}^{i}_{a}{\Tilde F}
~&=~  ({\cal U}{}_{3}){}^{i j} (\g^{\mu})_{a}{}^{b} \pa_{\mu} \Psi^{j}_{b}  ~~~, \\
 {\Tilde {\text D}}{}^{i}_{a}{\Tilde G} ~&=~ i ({\cal U}{}_{4}){}^{ij} (\g^{5}\g^{
 \mu})_{a}{}^{b} \pa_{\mu} \Psi^{j}_{b}  ~~~, \\
{\Tilde {\text D}}{}^{i}_{a}{\Tilde {\varphi}} ~&=~ ({\cal U}{}_{5}){}^{ij} {\Psi}^{j}_{a}  ~~~, \\
 {\Tilde {\text D}}{}^{i}_{a}{\Tilde B}_{\mu\nu}
~&=~ - i \tfrac{1}{4}  ({\cal U}{}_{6}){}^{ij}  ([\g_{\mu},\g_{\nu}])_{a}{}^{b} { {\Psi}}^{j}_{b}  ~~~, \\
 {\Tilde {\text D}}{}^{i}_{a}{\Tilde {\Psi}}^{j}_{b}
 ~&=~   \big[ i   ({\cal U}{}_{7}){}^{ij} (\g^{\mu})_{ab} \pa_{\mu} A -   ({\cal U
 }{}_{8}){}^{ij} (\g^{5}\g^{\mu})_{ab} \pa_{\mu} B ~~~, \\
& ~~~~~- i   ({\cal U}{}_{9}){}^{ij} C_{ab} F \big] +   ({\cal U}{}_{10}){}^{ij} (\g^{
5})_{ab} G  ~~~, \\
 & ~~~+  ({\cal U}{}_{11}){}^{ij} (\g^{5})_{ab} d + \tfrac{1}{4}  ({\cal U}{}_{12}){}^{ij}  
 ([\g^{\mu},\g^{\nu}])_{ab} (\pa_{\mu} A_{\nu} - \pa_{\nu} A_{\mu})
 ~~~,
\end{split}
\ee
on the fields in the 4D, $\cal N$ = 2 tensor supermultiplet.

We next seek solutions for the $(\mathcal{U}_n)^{ij}$ and $(\mathcal{V}_n)^{ij}$ matrices that lead 
to invariance of the Lagrangian $\mathcal{L}_{(4TV)} = \mathcal{L}_{(2VS)} + \mathcal{L
}_{(2TS)}$:
\begin{align}
{\tilde {\text D}}{}^{i}_{a} \mathcal{L}_{(4TV)} = 0 + \text{total derivative  ~~~.}
\end{align}
The solution to this condition is given by
\begin{align}
(\mathcal{U}_{1})^{ij} =&(\mathcal{V}_{10})^{ij} ~~~,~~~(\mathcal{U}_2)^{ij}=(\mathcal{
V}_7)^{ij} ~~~,~~~(\mathcal{U}_{3})^{ij}=(\mathcal{V}_8)^{ij}~~~,~~~(\mathcal{U}_{4}
)^{ij}=(\mathcal{V}_9)^{ij} ~~~,\cr
(\mathcal{U}_5)^{ij}=&(\mathcal{V}_{11})^{ij}~~~,~~~(\mathcal{U}_{6})^{ij} =(\mathcal{V
}_{12})^{ij}~~~,~~~(\mathcal{U}_7)^{ij}=(\mathcal{V}_1)^{ij}~~~,~~~(\mathcal{U}_8)^{ij}
=(\mathcal{V}_2)^{ij}~~~,\cr
(\mathcal{U}_9)^{ij}=&(\mathcal{V}_3)^{ij}~~~,~~~(\mathcal{U}_{10})^{ij}=(\mathcal{V
}_4)^{ij}~~~,~~~(\mathcal{U}_{11})^{ij}=(\mathcal{V}_6)^{ij}~~~,~~~
(\mathcal{U}_{12})^{ij}=(\mathcal{V}_5)^{ij}~~~.
\end{align}

For no choice of the $(\mathcal{V}_n)^{ij}$ matrices does the algebra close for $\{ \text D^i_a, 
\Tilde {\text D}^j_b \}$ without central charges. This has been verified via \emph{Mathematica} code that we have made available open-source at the \href{https://hepthools.github.io/Data/}{HEPTHools Data Repository}.  In contrast, there are 327,680 ways to close the $\{ \Tilde{\text D}^i_a,  
\Tilde{\text D}^j_b \}$ portion of the algebra, up to gauge transformations. One of these choices 
is as follows (the $ij$ indices are suppressed below, and also below $\text I$ refers to the $ 2\times 
2$ identity 
matrix):
\begin{align}\label{e:V}
\mathcal{V}_n = \left\{i \s _2,i \s _2,i \s_2,\s _1,-i \text{I},-\s _3,i
\s _2,i \s _2,i \s_2,-\s _1,\s _3,-i
\text{I}\right\} ~~~.
\end{align}
For this choice, the $\{ \Tilde{\text D}^i_a,  \Tilde{\text D}^j_b \}$ portion of the algebra takes the form
\begin{align}
 \{\Tilde{\text{D}}^{i}_{a},\Tilde{\text{D}}^{j}_{b}\} \chi =& i 2 \delta^{ij} (\g^{\mu})_{ab} 
 \pa_{\mu} \chi  ~~~, \\ 
 \{\Tilde{\text{D}}^{i}_{a},\Tilde{\text{D}}^{j}_{b}\} A_{\mu} ~&=~ i 2 \delta^{ij} (\g^{\nu})_{
 ab} F_{\nu\mu} - i 2 (\s^{2})^{ij} \pa_{\mu} \left[ i  C_{ab} A - (\g^{5})_{ab}  B  \right] ~~~, \\
\begin{split}
\{\Tilde{\text{D}}^{i}_{a},\Tilde{\text{D}}^{j}_{b}\} {\Tilde B}_{\mu\nu} =& i 2 \delta^{ij} 
(\g^{\alpha})_{ab} \partial_\a \Tilde{B}_{\mu\nu} - \pa_{[\mu} \Tilde{\Tilde{\Lambda
}}^{ij}_{\nu]ab}  ~~~, 
\end{split}
\end{align}
where $\chi = \{A,B,F,G,d,\Psi^{k}_{c},{\Tilde A},{\Tilde B},{\Tilde F},{\Tilde G},{\Tilde 
{\varphi}},{\Tilde {\Psi}}^{k}_{c}\} $
and 
\begin{align}
 \Tilde{\Tilde{\Lambda}}^{ij}_{\nu ab} = & 2i \delta^{ij} (\g^\alpha)_{ab} B_{\alpha 
 \nu} + i (\s^1)^{ij} (\g_\nu)_{ab} \Tilde{A} +i (\s^2)^{ij} (\g^5 
 \g_\nu)_{ab} \Tilde{B} - i (\s^3)^{ij} (\g_\nu)_{ab} \Tilde{\varphi}~~~.
\end{align}
The cross terms $\{ {\text D}^i_a,  \Tilde{\text D}^j_b \}$ take the form
\begin{align}
\{ {\text D}^i_a,  \Tilde{\text D}^j_b  \} A =& -2 i (\s^1)^{ij} (\g^\mu)_{ab} \pa_\mu 
\Tilde{A} + 2 i (\s^3)^{ij} (\g^\mu)_{ab} \pa_\mu \Tilde{\varphi} ~~~, \\
\{ {\text D}^i_a,  \Tilde{\text D}^j_b \} B =& 2 (\s^2)^{ij} (\g^\mu)_{ab} \pa_\mu 
\Tilde{B} + 2 i  \delta^{ij} \epsilon^{\mu\nu\alpha\beta}(\g_{\mu})_{ab} \pa_\nu 
\Tilde{B}_{\alpha\beta} ~~~, \\
\{ {\text D}^i_a,  \Tilde{\text D}^j_b \} F =& 2 (\s^2)^{ij} (\g^\mu)_{ab} \pa_\mu 
\Tilde{F} + i \delta^{ij} ([\g^\alpha, \g^\beta])_{ab} \square \Tilde{B}_{\alpha\beta}  
- 2 i \delta^{ij} ([\g^\alpha, \g^\mu])_{ab} \pa_\mu \pa^\beta \Tilde{B}_{
\alpha\beta} ~~~, \\
\{ {\text D}^i_a,  \Tilde{\text D}^j_b \} G =& 2 i (\s^2)^{ij} (\g^5)_{ab} \square 
\Tilde{A} + 2 i (\s^1)^{ij} C_{ab} \square \Tilde{B} + 2 (\s^1)^{ij} (\g^5 
\g^\mu)_{ab} \pa_\mu \Tilde{F} \cr
&- (\s^3)^{ij} (\g^5 [\g^\alpha, \g^\beta])_{ab} \square \Tilde{
B}_{\alpha\beta} + 2 (\s^3)^{ij} (\g^5 [\g^\alpha, \g^\mu])_{ab
} \pa_\mu \pa^\beta \Tilde{B}_{\alpha\beta} ~~~, \\
\{ {\text D}^i_a,  \Tilde{\text D}^j_b \} d =& -2 i  (\s^3)^{ij} C_{ab} \square \Tilde{B} - 
2 (\s^3)^{ij} (\g^5 \g^\mu)_{ab} \pa_\mu \Tilde{F} + 2 i (\s^2
)^{ij} (\g^5)_{ab} \square \Tilde{\varphi} \cr
&- (\s^1)^{ij} (\g^5 [\g^\alpha, \g^\beta])_{ab} \square \Tilde{
B}_{\alpha\beta} + 2 (\s^1)^{ij} (\g^5 [\g^\alpha, \g^\mu])_{ab
} \pa_\mu \pa^\beta \Tilde{B}_{\alpha\beta} ~~~, \\
\{ {\text D}^i_a,  \Tilde{\text D}^j_b  \} \Tilde{A} =& -2 i (\s^1)^{ij} (\g^\mu)_{ab
} \pa_\mu A  - 2 i (\s^2)^{ij} (\g^5)_{ab} G  ~~~, \\
\{ {\text D}^i_a,  \Tilde{\text D}^j_b \} \Tilde{B} =& 2 (\s^2)^{ij} (\g^\mu)_{ab
} \pa_\mu B - 2 i  (\s^1)^{ij} C_{ab} G + 2 i (\s^3)^{ij} C_{ab} d + \delta^{
ij} (\g^5 [\g^\mu, \g^\nu])_{ab} \pa_\mu A_\nu  ~~~, \\
\{ {\text D}^i_a,  \Tilde{\text D}^j_b \} \Tilde{F} =& 2 (\s^2)^{ij} (\g^\mu)_{
ab} \pa_\mu F+ 2 (\s^1)^{ij} (\g^5 \g^\mu)_{ab} \pa_\mu 
G - 2 (\s^3)^{ij} (\g^5 \g^\mu)_{ab} \pa_\mu d  \cr
&-2 i \delta^{ij} (\g^\mu)_{ab} \square A_\mu + 2 i \delta^{ij} (\g^\mu)_{
ab} \pa_\mu \pa^\nu A_\nu ~~~, \\
\{ {\text D}^i_a,  \Tilde{\text D}^j_b \} \Tilde{G} =&  0  ~~~, \\
\{ {\text D}^i_a,  \Tilde{\text D}^j_b  \} \Tilde{\varphi}=& 2 i (\s^3)^{ij} (\g^\mu
)_{ab} \pa_\mu A - 2 i (\s^2)^{ij} (\g^5)_{ab} d  ~~~, \\
\{\text{D}^{i}_{a},\Tilde{\text{D}}^{j}_{b}\} A_{\mu} ~&=-\delta^{ij} (\g^5 [\g_\mu, 
\g_\nu])_{ab} \pa^\nu \Tilde{B} + 2 i \delta^{ij} (\g_\mu)_{ab} \Tilde{F} - 
2 (\s^2)^{ij} ([ \g^\alpha, \g^\beta])_{ab} \pa_\alpha \Tilde{B}_{
\beta\mu} - \pa_\mu \Tilde{\lambda}^{ij}_{ab}  ~~~, \\
\{\text{D}^{i}_{a},\Tilde{\text{D}}^{j}_{b}\} \Tilde{B}_{\mu\nu} ~&= - \frac{1}{2} \delta^{ij} 
(\g^5\g^\alpha[\g_\mu, \g_\nu])_{ab}\pa_\alpha B - i \frac{1}{2}
\delta^{ij} ( [\g_\mu, \g_\nu])_{ab} F + \frac{1}{2} (\s^3)^{ij} (\g^5 
[\g_\mu, \g_\nu])_{ab} G   \cr
&{~~~} + \frac{1}{2} (\s^1)^{ij} (\g^5 [\g_\mu , \g_\nu])_{ab} d  -\frac{1}{2} 
(\s^2)^{ij} ([\g_\alpha, \g_{[\mu}])_{|ab|} \pa^\alpha A_{\nu]}- 
\pa_{[\mu} \Tilde{\Lambda}^{ij}_{ \nu ] ab} ~~~,  \\
 \{\text{D}^{i}_{a},\Tilde{\text{D}}^{j}_{b}\} \Psi^k_c ~&= 2 i Z_1^{ijkl} (\g^\mu)_{ab} 
 \pa_\mu \Tilde{\Psi}^l_c + 2 i Z_2^{ijkl} ([\g^\mu, \g^\nu])_{ab}(\g_\nu
)_c{}^d \pa_\mu \Tilde{\Psi}^l_d      \cr
&{~~~} + 2 i Z_3^{ijkl} (\g_\nu )_{ab}([\g^\nu, \g^\mu])_c{}^d \pa_\mu 
\Tilde{\Psi}^l_d +2 i Z_4^{ijkl} (\g^5[\g^\mu, \g^\nu])_{ab}(\g^5
\g_\nu)_c{}^d \pa_\mu \Tilde{\Psi}^l_d   \cr
 &{~~~} +2 i Z_5^{ijkl} (\g^5 \g^\mu)_{ab} (\g^5)_c{}^d \pa_\mu \Tilde{
 \Psi}^l_d + 2 i Z_6^{ijkl} (\g^5\g_\nu )_{ab}(\g^5[\g^\nu, 
 \g^\mu])_c{}^d \pa_\mu \Tilde{\Psi}^l_d \cr
 &{~~~} + 2 i Z_7^{ijkl} (\g^5)_{ab} (\g^5 \g^\mu)_c{}^d \pa_\mu 
 \Tilde{\Psi}_d^l + 2 i Z_8^{ijkl} C_{ab} (\g^\mu)_c{}^d \pa_\mu \Tilde{\Psi}_d^l  
 ~~~, \\
\{\text{D}^{i}_{a},\Tilde{\text{D}}^{j}_{b}\} \Tilde{\Psi}^k_c ~&= 2 i \Tilde{Z}_1^{ijkl} 
(\g^\mu)_{ab} \pa_\mu \Psi^l_c + 2 i \Tilde{Z}_2^{ijkl} ([\g^\mu, 
\g^\nu])_{ab}(\g_\nu)_c{}^d \pa_\mu \Psi^l_d   \cr
&{~~~} + 2 i \Tilde{Z}_3^{ijkl} (\g_\nu )_{ab}([\g^\nu, \g^\mu])_c{}^d \pa_\mu 
\Psi^l_d +2 i \Tilde{Z}_4^{ijkl} (\g^5[\g^\mu, \g^\nu])_{ab}(\g^5
\g_\nu)_c{}^d \pa_\mu \Psi^l_d       \cr
&{~~~} +2 i \Tilde{Z}_5^{ijkl} (\g^5 \g^\mu)_{ab} (\g^5)_c{}^d \pa_\mu 
\Psi^l_d + 2 i \Tilde{Z}_6^{ijkl} (\g^5\g_\nu )_{ab}(\g^5[\g^\nu, 
\g^\mu])_c{}^d \pa_\mu \Psi^l_d       \cr
&{~~~} + 2 i \Tilde{Z}_7^{ijkl} (\g^5)_{ab} (\g^5 \g^\mu)_c{}^d \pa_\mu 
\Psi_d^l + 2 i \Tilde{Z}_8^{ijkl} C_{ab} (\g^\mu)_c{}^d \pa_\mu \Psi_d^l   ~~~,
\end{align}
where the parameters that associated with the commutators of the
gauge fields above are given by
\begin{align}
\Tilde{\lambda}^{ij}_{ab} = & (\s^2)^{ij} ([\g^\alpha, \g^\beta])_{ab} 
\Tilde{B}_{\alpha\beta} + 2 i (\s^3)^{ij} C_{ab} \Tilde{A} + 2 i (\s^1)^{ij} C_{
ab} \Tilde{\varphi}    ~~~,\\
\Tilde{\Lambda}^{ij}_{\nu ab} = & - \frac{1}{2} (\s^2)^{ij} [\g_\nu , \g_\alpha 
]_{ab} A^\alpha - i \delta^{ij} (\g_\nu)_{ab} A - \delta^{ij} (\g^5 \g_\nu
)_{ab} B   ~~~.
\end{align}
In a similar manner, the $Z$-factors and $\tilde Z$-factors that appear in the commutators
associated with the spinors are defined by
\begin{align}
Z_1^{\text{ijkl}}=&-\frac{3}{2} (\s^1) ^{\text{ij}} \left(\s^3\right)^{\text{kl}}-\frac{1}{4} 
(\s^1) ^{\text{ik}} \left(\s^3\right)^{\text{jl}}+\frac{3}{4} \left(\s ^3\right)^{\text{il}} 
(\s^1)^{\text{jk}}+(\s^1) ^{\text{il}} \left(\s^3\right)^{\text{jk}}   ~~~,    \cr
Z_2^{\text{ijkl}}=&-\frac{1}{4} (\s^1) ^{\text{ij}} \left(\s ^3\right)^{\text{kl}}+\frac{1}{8} 
(\s^1) ^{\text{ik}} \left(\s^3\right)^{\text{jl}}+\frac{1}{8} \left(\s ^3\right)^{\text{il}} 
(\s^1)^{\text{jk}}    ~~~,    \cr
Z_3^{\text{ijkl}}=&\frac{1}{8} \left(\s ^3\right)^{\text{il}}(\s^1) ^{\text{jk}}-\frac{1}{8} 
(\s^1) ^{\text{ik}} \left(\s^3\right)^{\text{jl}}     ~~~, \cr
Z_4^{\text{ijkl}}=&\frac{1}{8} (\s^1) ^{\text{ik}} \left(\s ^3\right)^{\text{jl}}-\frac{1}{8} 
\left(\s ^3\right)^{\text{il}} (\s^1) ^{\text{jk}}   ~~~, \cr
Z_5^{\text{ijkl}}=&-\frac{1}{2} (\s^1) ^{\text{ij}} \left(\s ^3\right)^{\text{kl}}-\frac{3}{4} 
(\s^1) ^{\text{ik}} \left(\s^3\right)^{\text{jl}}+\frac{3}{4} \left(\s ^3\right)^{\text{il}} 
(\s^1)^{\text{jk}}+\frac{1}{2} (\s^1) ^{\text{il}} \left(\s^3\right)^{\text{jk}} ~~~,  \cr
Z_6^{\text{ijkl}}=&-\frac{3}{8} (\s^1) ^{\text{ik}} \left(\s ^3\right)^{\text{jl}}+\frac{1}{8} 
\left(\s ^3\right)^{\text{il}} (\s^1) ^{\text{jk}}+\frac{1}{4} (\s^1) ^{\text{il}} \left(
\s^3\right)^{\text{jk}}   ~~~, \cr
Z_7^{\text{ijkl}}=&\frac{1}{4} (\s^1) ^{\text{ik}} \left(\s ^3\right)^{\text{jl}}+\frac{1}{4} 
\left(\s ^3\right)^{\text{il}} (\s^1) ^{\text{jk}}-\frac{1}{2} (\s^1) ^{\text{il}} \left(
\s^3\right)^{\text{jk}}    ~~~, \cr
Z_8^{\text{ijkl}}=&-\frac{1}{2} (\s^1) ^{\text{ij}} \left(\s ^3\right)^{\text{kl}}-\frac{3}{4} 
(\s^1) ^{\text{ik}} \left(\s^3\right)^{\text{jl}}+\frac{3}{4} \left(\s ^3\right)^{\text{il}} 
(\s^1)^{\text{jk}}+\frac{1}{2} (\s^1) ^{\text{il}} \left(\s^3\right)^{\text{jk}}  ~~~, \cr
\Tilde{Z}_1^{\text{ijkl}}=&-\frac{3}{2} (\s^1) ^{\text{ij}} \left(\s^3\right)^{\text{kl}}+
\frac{1}{4} (\s^1) ^{\text{ik}} \left(\s^3\right)^{\text{jl}}+\frac{3}{4} \left(\s ^3
\right)^{\text{il}} (\s^1)^{\text{jk}}+\frac{1}{2} (\s^1) ^{\text{il}} \left(\s^3\right)^{
\text{jk}}   ~~~, \cr
\Tilde{Z}_2^{\text{ijkl}}=&-\frac{1}{4} (\s^1) ^{\text{ij}} \left(\s ^3\right)^{\text{kl}}-\frac{1}{8} 
(\s^1) ^{\text{ik}} \left(\s^3\right)^{\text{jl}}+\frac{1}{8} \left(\s ^3\right)^{\text{il}} 
(\s^1)^{\text{jk}}+\frac{1}{4} (\s^1) ^{\text{il}} \left(\s^3\right)^{\text{jk}}\cr
\Tilde{Z}_3^{\text{ijkl}}=&\frac{1}{8} \left(\s^3\right)^{\text{il}} (\s^1) ^{\text{jk}}-\frac{1}{8} 
(\s^1) ^{\text{ik}} \left(\s ^3\right)^{\text{jl}}  ~~~, \cr
\Tilde{Z}_4^{\text{ijkl}}=&\frac{1}{8} (\s^1)^{\text{ik}} \left(\s ^3\right)^{\text{jl}}-\frac{1}{8} 
\left(\s^3\right)^{\text{il}} (\s^1) ^{\text{jk}}   ~~~, \cr
\Tilde{Z}_5^{\text{ijkl}}=&\frac{1}{2} (\s^1) ^{\text{ij}} \left(\s ^3\right)^{\text{kl}}+
\frac{1}{4} (\s^1)^{\text{ik}} \left(\s ^3\right)^{\text{jl}}-\frac{3}{4} \left(\s^3\right)^{
\text{il}} (\s^1) ^{\text{jk}}   ~~~, \cr
\Tilde{Z}_6^{\text{ijkl}}=&-\frac{1}{8} (\s^1) ^{\text{ik}} \left(\s ^3\right)^{\text{jl}}-\frac{1}{8} 
\left(\s^3\right)^{\text{il}} (\s^1) ^{\text{jk}}+\frac{1}{4} (\s^1) ^{\text{il}} \left(\s ^3
\right)^{\text{jk}}   ~~~, \cr
\Tilde{Z}_7^{\text{ijkl}}=&\frac{3}{4} (\s^1)^{\text{ik}} \left(\s ^3\right)^{\text{jl}}-\frac{1}{4} 
\left(\s^3\right)^{\text{il}} (\s^1) ^{\text{jk}}-\frac{1}{2} (\s^1) ^{\text{il}} \left(\s^3
\right)^{\text{jk}}   ~~~ , \cr
\Tilde{Z}_8^{\text{ijkl}}=&\frac{1}{2} (\s^1)^{\text{ij}} \left(\s ^3\right)^{\text{kl}}+\frac{1}{4} 
(\s^1) ^{\text{ik}} \left(\s ^3\right)^{\text{jl}}-\frac{3}{4} \left(\s ^3\right)^{\text{il}}
(\s^1) ^{\text{jk}}   ~~~.
\end{align}

To summarize the results seen here, the 4D, $\cal N$ = 4 vector-tensor supermultiplet can
be realized in terms of one 4D, $\cal N$ = 2 vector-supermultiplet and one 4D, $\cal N$ = 2 
tensor-supermultiplet.  The form of the super algebra of its four supercharges ${{\text 
D}}{}^{i}_{a}$ is and ${\tilde {\text D}}{}^{i}_{a}$ is
\be \eqalign{
 \{{\text{D}}^{i}_{a},{\text{D}}^{j}_{b}\}  ~&=~
 \{\Tilde{\text{D}}^{i}_{a},\Tilde{\text{D}}^{j}_{b}\}  ~=~ i 2 \delta^{ij} (\g^{\mu})_{ab} 
 \pa_{\mu} ~~~, \cr
\{{\text{D}}^{i}_{a}, \Tilde{\text{D}}^{j}_{b}  \}  ~&=~ {\rm {central ~charge}} ~~~,
}\ee
uniformly on all of the component fields.

\section{Hat Derivatives Transformation Laws}\label{s:cv}

As the spins of the states in 2D, $\cal N$ = 2 vector-tensor supermultiplet are the same as 
in the Wess-Fayet 4D, $\cal N$ = 2 supermultiplet  \cite{wF1,wF2}, there must be a formulation 
of the latter that is similar to the construction in chapter two.  We thus introduce an
ansatz of the form
\be\label{e:Wansatz}
\begin{split}
{\hat {\text D}}{}^{i}_{a} A 
~&=~   ({\cal W}{}_{1}){}^{i j} {\hat {\Psi}}^{j}_{a}  ~~~, \\
{\hat {\text D}}{}^{i}_{a} B 
~&=~ i  ({\cal W}{}_{2}){}^{i j}(\g^{5})_{a}{}^{b} {\hat {\Psi}}^{j}_{b}  ~~~, \\
{\hat {\text D}}{}^{i}_{a} F  ~&=~  ({\cal W}{}_{3}){}^{i j} (\g^{\mu})_{a}{}^{b} \pa_{\mu} 
{\hat {\Psi}}^{i}_{b}  ~~~, \\
{\hat {\text D}}{}^{i}_{a} G  ~&=~ i  ({\cal W}{}_{4}){}^{i j} (\g^{5}\g^{\mu})_{a}{}^{b} 
\pa_{\mu} {\hat {\Psi}}^{j}_{b}  ~~~, \\
{\hat {\text D}}{}^{i}_{a} A_{\mu} ~&=~ ({\cal W}{}_{5}){}^{ij} (\g_{\mu})_{a}{}^{b} 
{\hat \Psi}{}^{j}_{b}  ~~~, \\
 {\hat {\text D}}{}^{i}_{a} d  ~&=~ i ({\cal W}{}_{6}){}^{i j} (\g^{5}\g^{\mu})_{a}{}^{b} 
 \pa_{\mu} {\hat {\Psi}} {}^{j}_{b}  ~~~, \\
 {\hat {\text D}}{}^{i}_{a} \Psi^{j}_{b} ~&=~  i({\cal W}_7)^{ij}(\g^{\mu})_{ab}\pa_{
 \mu}\hat{A}- i({\cal W}_8)^{ij}C_{ab}\hat{F}  -({\cal W}_{9})^{ij}(\g^5\g^{\mu})_{ab}
 \pa_{\mu}\hat{B} + ({\cal W}_{10})^{ij}(\g^5_{ab}) \hat{G} ~~~, \\
 &{~~~~} +i({\cal W}_{11})^{ij}(\g^{\mu})_{ab}\pa_{\mu}\breve{A}-i({\cal W}_{12})^{ij}C_{ab}
 \breve{F} -({\cal W}_{13})^{ij}(\g^5\g^{\mu})_{ab}\pa_{\mu}\breve{B} + ({\cal W}_{
 14})^{ij}(\g^5_{ab})\breve{G}
\end{split}
\ee
\begin{align}\label{e:Xansatz}
\begin{split}
  \hat{\rm D}_a^i \hat{A} &= ({\cal X}_1)^{ij}\Psi_a^j, ~~~, \\
  \hat{\rm D}_a^i \hat{B} &= i ({\cal X}_2)^{ij}(\g^5)_a^{~b} \Psi_b^j, ~~~, \\
  \hat{\rm D}_a^i \hat{F} &= ({\cal X}_3)^{ij}(\g^{\mu})_a^{~b} \pa_{\mu}\Psi_b^j,~~~, \\
  \hat{\rm D}_a^i \hat{G} &= i ({\cal X}_4)^{ij}(\g^5\g^\mu)_a^{~b}\pa_\mu \Psi_b^j,~~~, \\
  \hat{\rm D}_a^i \breve{A} &= ({\cal X}_5)^{ij}\Psi_a^j, ~~~, \\
  \hat{\rm D}_a^i \breve{B} &= i({\cal X}_6)^{ij} (\g^5)_a^{~b} \Psi_b^j,~~~, \\
  \hat{\rm D}_a^i \breve{F} &= ({\cal X}_7)^{ij}(\g^{\mu})_a^{~b} \pa_{\mu}\Psi_b^j,~~~, \\
  \hat{\rm D}_a^i \breve{G} &= i({\cal X}_8)^{ij} (\g^5\g^\mu)_a^{~b}\pa_\mu 
  \Psi_b^j,~~~, \\
  \hat{\rm D}_a^i \hat{\Psi}_b^j &= i  ({\cal X}_9)^{ij}  (\g^{\mu})_{ab} \pa_{\mu} A -  
  ({\cal X}_{10})^{ij} (\g^{5}\g^{\mu})_{ab} \pa_{\mu} B - i ({\cal X}_{11})^{ij}  
  C_{ab} F  + ({\cal X}_{12})^{ij} (\g^{5})_{ab} G  ~~~, \\
 &{~~~}~  + ({\cal X}_{13})^{ij} (\g^{5})_{ab} d - \tfrac{1}{4} i ({\cal X}_{14})^{ij} ([\g^{\mu},
 \g^{\nu}])_{ab} (\pa_{\mu} A_{\nu} - \pa_{\nu} A_{\mu}) ~~~.
 \end{split}
\end{align}
for the fields of the 4D, $\cal N$ = 2 vector supermultiplet combined with the fields of
the 4D, $\cal N$ = 2 W-F supermultiplet.

We next seek solutions for the $(\mathcal{W}_n)^{ij}$ and $(\mathcal{X}_n)^{ij}$ that lead to 
invariance of the Lagrangian $\mathcal{L}_{(4CV)} = \mathcal{L}_{(2VS)} + \mathcal{L}_{(2CC)}$:
\begin{align}
	 {\hat {\text D}}{}^{i}_{a} \mathcal{L}_{(4CV)} = 0 + \text{total derivative.}
\end{align}
This solution is
\begin{align}\label{e:XW}
(\mathcal{X}_{1})^{ij} =&(\mathcal{W}_{7})^{ij} ~~~,~~~(\mathcal{X}_2)^{ij}=(\mathcal{W}_9)^{ij} 
~~~,~~~(\mathcal{X}_{3})^{ij}=(\mathcal{W}_8)^{ij}~~~,~~~(\mathcal{X}_{4})^{ij}=(\mathcal{
W}_{10})^{ij} ~~~,\cr
(\mathcal{X}_5)^{ij}=&(\mathcal{W}_{11})^{ij}~~~,~~~(\mathcal{X}_{6})^{ij}=(\mathcal{W}_{13
})^{ij}~~~,~~~(\mathcal{X}_7)^{ij}=(\mathcal{W}_{12})^{ij}~~~,~~~(\mathcal{X}_8)^{ij}=(\mathcal{
W}_{14})^{ij}~~~,\cr
(\mathcal{X}_9)^{ij}=&(\mathcal{W}_1)^{ij}~~~,~~~(\mathcal{X}_{10})^{ij}=(\mathcal{W}_2)^{
ij}~~~,~~~(\mathcal{X}_{11})^{ij}=(\mathcal{W}_3)^{ij}~~~,~~~
(\mathcal{X}_{12})^{ij}=(\mathcal{W}_4)^{ij}~~~,\cr
(\mathcal{X}_{13})^{ij}=&(\mathcal{W}_6)^{ij}~~~,~~~(\mathcal{X}_{14})^{ij}=(\mathcal{W}_5)^{ij}  ~~~.
\end{align}

With the use of open-source \emph{Mathematica} code that can be found at the \href{https://hepthools.github.io/Data/}{HEPTHools Data Repository}, we have found that even without imposing the above Lagrangian constraints, 
for no choice of $(\mathcal{W}_n)^{ij}$ does the algebra $ \{ \hat{\rm D}_a^i ,  \hat{\rm D}_b^j \}$ or  
$ \{ {\rm D}_a^i ,  \hat{\rm D}_b^j \}$ close on the fields of the chiral-chiral W-F hypermultiplet. That 
is for any possible choice of $(\mathcal{W}_n)^{ij}$, we necessarily have
\begin{align}
\{ \hat{\rm D}_a^i ,  \hat{\rm D}_b^j \} X \ne \delta^{ij} ( \g^\m)_{ab} \pa_\mu X ~~~,~~~\{ {\rm D}_a^i , 
 \hat{\rm D}_b^j \} X \ne 0  ~~~,
\end{align}
where $X$ is at least one of the fields in the list
\begin{align}
X \subset \hat{A}, \breve{A}, \hat{B}, \breve{B}, \hat{F}, \breve{F}, \hat{G}, \breve{G}, \hat{\Psi}^i_a~~~.
\end{align}

As such, we chose the $(\mathcal{W}_n)^{ij}$ such that the $\hat{\rm D}_a^i$ transformation laws 
parallel those of the ${\rm D}_a^i$ transformation laws, i.e., ${\rm D}_a^i \to \hat{\rm D}_a^i $ and 
$\Psi_a^i \to \hat{\Psi}_a^i$:
\begin{align}\label{e:W}
\mathcal{W}_{n} =& \{~ {\rm I}~,~{\rm I}~,~{\rm I}~,~\s^3~,~i\s^2~,~\s^1~,~\s^3~,~\s^3~,~{\rm I}~,~
{\rm I}~,~\s^1~,~\s^1~,~i\s^2~,~i\s^2~    \}~~~,
\end{align}
which upon enforcing Eq.~(\ref{e:XW}) demands
\begin{align}\label{e:X}
\mathcal{X}_{n} =& \{~\s^3~,~{\rm I}~,~\s^3~,~{\rm I}~,~\s^1~,~i\s^2~,~\s^1~,~i\s^2~,~{\rm I}~,~
{\rm I}~,~{\rm I}~,~\s^3~,~\s^1~,~i\s^2~\}  ~~~.
\end{align}

With the above choices for $(\mathcal{W}_n)^{ij}$ and $(\mathcal{X}_n)^{ij}$, the  transformation 
laws satisfy the following $\{\hat{\text{D}}^{i}_{a},\hat{\text{D}}^{j}_{b}\}$ algebra on the fields of 
the $\mathcal{N}=2$ vector multiplet
\begin{align}
 \{\hat{\text{D}}^{i}_{a},\hat{\text{D}}^{j}_{b}\} \chi ~&=~ i 2 \delta^{ij} (\g^{\mu})_{ab} \pa_{\mu} \chi ~~~, \\ 
 \{\hat{\text{D}}^{i}_{a},\hat{\text{D}}^{j}_{b}\} A_{\mu} ~&=~ i 2 \delta^{ij} (\g^{\nu})_{ab} F_{\nu\mu} 
 + i \, 2(\s^{2})^{ij} \,  \pa_{\mu}  \left[ i C_{ab}A - (\g^{5})_{ab} B  \right]  ~~~,
\end{align}
where
\be
 \chi \in \{A,B,F,G,d,\Psi^{k}_{c}\}~~~.
\ee

For the fields of the $\mathcal{N}=2$ W-F hypermultiplet, the $\{\hat{\text{D}}^{i}_{a},\hat{\text{D}}^{j}_{b}\}$ 
algebra is
\begin{align}\label{e:CChathatAlgebra}
\begin{split}
\left\{\hat{\rm D}^i_a,\hat{\rm D}^j_b\right\}\hat{A}=&\delta^{ij}2i\left(\g^\mu\right)_{ab}\pa_\mu \hat{A}+i\left(
\s^2\right)^{ij}2iC_{ab}\breve{F}~~, \\ 
\left\{\hat{\rm D}^i_a,\hat{\rm D}^j_b\right\}\breve{A}=&\delta^{ij}2i\left(\g^\mu\right)_{ab}\pa_\mu\breve{A}-i\left(
\s^2\right)^{ij}2iC_{ab}\hat{F}~~, \\ 
\left\{\hat{\rm D}^i_a,\hat{\rm D}^j_b\right\}\hat{B}=&\delta^{ij}2i\left(\g^\mu\right)_{ab}\pa_\mu \hat{B}+i\left(\s^2
\right)^{ij}2iC_{ab}\breve{G}~~, \\ 
\left\{\hat{\rm D}^i_a,\hat{\rm D}^j_b\right\}\breve{B}=&\delta^{ij}2i\left(\g^\mu\right)_{ab}\pa_\mu\breve{B}-i\left(
\s^2\right)^{ij}2iC_{ab} \hat{G}~~, \\ 
\left\{\hat{\rm D}^i_a,\hat{\rm D}^j_b\right\}\hat{F}=&\delta^{ij}2i\left(\g^\mu\right)_{ab}\pa_\mu \hat{F}+i\left(\s^2
\right)^{ij}2iC_{ab}\square\breve{A}~~, \\ 
\left\{\hat{\rm D}^i_a,\hat{\rm D}^j_b\right\}\breve{F}=&\delta^{ij}2i\left(\g^\mu\right)_{ab}\pa_\mu\breve{F}-i\left(
\s^2\right)^{ij}2iC_{ab}\square \hat{A}~~, \\ 
\left\{\hat{\rm D}^i_a,\hat{\rm D}^j_b\right\}\hat{G}=&\delta^{ij}2i\left(\g^\mu\right)_{ab}\pa_\mu \hat{G}+i\left(
\s^2\right)^{ij}2iC_{ab}\square\breve{B}~~, \\ 
\left\{\hat{\rm D}^i_a,\hat{\rm D}^j_b\right\}\breve{G}=&\delta^{ij}2i\left(\g^\mu\right)_{ab}\pa_\mu\breve{G}-i\left(
\s^2\right)^{ij}2iC_{ab}\square \hat{B}~~, \\ 
\left\{\hat{\rm D}^i_a,\hat{\rm D}^j_b\right\}\hat{\Psi}^k_c=&\delta^{ij}2i\left(\g^\mu\right)_{ab}\pa_\mu \hat{\Psi
}^k_c-\left(\s^2\right)^{ij}\left(\s^2\right)^{kr}
2iC_{ab}\left(\g^\mu\right)_{c}^{~d}\pa_\mu\hat{\Psi}^r_d ~~~.
\end{split}
\end{align}
For the cross terms $\{\text{D}^{i}_{a},\hat{\text{D}}^{j}_{b}\}$, we have the following algebra for the bosons
\begin{align}
\left\{\text{D}^{i}_{a},\hat{\text{D}}^{j}_{b}\right\}A=&2 i (\s^3)^{ij} (\g^\m)_{ab} \pa_\m \hat{A} + 2 i (\s^1)^{ij} 
(\g^\m)_{ab} \pa_\m \breve{A} - 2 i (\s^2)^{ij} (\g^5\g^\m)_{ab} \pa_\m\breve{B} \cr
&+ 2 i (\s^2)^{ij}(\g^5)_{ab} \breve{G} ~~, \\ 
\left\{\text{D}^{i}_{a},\hat{\text{D}}^{j}_{b}\right\}B =& 2 i \delta^{ij} (\g^\m)_{ab} \pa_\m \hat{B} - 2 (\s^2)^{ij}
C_{ab} \breve{G} ~~, \\ 
\left\{\text{D}^{i}_{a},\hat{\text{D}}^{j}_{b}\right\} F = & 2 i (\s^3)^{ij}(\g^\m)_{ab} \pa_\m \hat{F} + 2 i (\s^2
)^{ij} (\g^5)_{ab} \square \breve{B} \cr
&+ 2 i (\s^1)^{ij} (\g^\m)_{ab} \pa_\m \breve{F} - 2 i (\s^2)^{ij} (\g^5\g^\m)_{ab}\pa_\m \breve{G} ~~, \\ 
\left\{\text{D}^{i}_{a},\hat{\text{D}}^{j}_{b}\right\} G = & 2 i (\s^3)^{ij} (\g^\m)_{ab} \pa_\m \hat{G} - 2 i (\s^2
)^{ij} (\g^5)_{ab} \square \breve{A}  \cr
&- 2 i (\s^2)^{ij} (\g^5\g^\m)_{ab} \pa_\m \breve{F} - 2 i (\s^1)^{ij}(\g^\m)_{ab} \pa_\m \breve{G} ~~, \\ 
\left\{\text{D}^{i}_{a},\hat{\text{D}}^{j}_{b}\right\} \rd =& 2 i (\s^2)^{ij}(\g^5)_{ab} \square \hat{A} + 2 i (
\s^2)^{ij} (\g^5\g^\m)_{ab} \pa_\m \hat{F} \cr
& + 2 i (\s^1)^{ij} (\g^\m)_{ab} \pa_\m \hat{G} + 2 i (\s^3)^{ij}(\g^\m)_{ab} \pa_\m \breve{G} ~~, \\ 
\left\{\text{D}^{i}_{a},\hat{\text{D}}^{j}_{b}\right\} A_\m =& - i (\s^1)^{ij}[\g_\m, \g_\n]_{ab} \pa^\n \hat{A} 
- 2 i (\s^1)^{ij}(\g_\m)_{ab} \hat{F} + 2 i (\s^2)^{ij} (\g^5\g_\m)_{ab} \hat{G} \cr
&+ i (\s^3)^{ij} [\g_\m, \g_\n]_{ab} \pa^\n \breve{A} - \delta^{ij} (\g^5[\g_\m, \g_\n])_{ab} \pa^\n \breve{B} 
+ 2 i (\s^3)^{ij}(\g_\m)_{ab} \breve{F} - \pa_\m \hat{\lambda}^{ij}_{ab} ~~, \\ 
\left\{\text{D}^{i}_{a},\hat{\text{D}}^{j}_{b}\right\} \hat{A} = & 2 i (\s^3)^{ij} (\g^\m)_{ab} \pa_\m A - 2 i 
(\s^2)^{ij}(\g^5)_{ab} \rd + i \fracm 12 (\s^1)^{ij} [\g^\m, \g^\n]_{ab} F_{\m\n} ~~, \\ 
\left\{\text{D}^{i}_{a},\hat{\text{D}}^{j}_{b}\right\} \breve{A} = & 2 i (\s^1)^{ij} (\g^\m)_{ab} \pa_\m A 
+ 2 i (\s^2)^{ij}(\g^5)_{ab} G - i \fracm 12 (\s^3)^{ij} [\g^\m, \g^\n]_{ab} F_{\m\n} ~~, \\ 
\left\{\text{D}^{i}_{a},\hat{\text{D}}^{j}_{b}\right\} \hat{B} =& 2 i \delta^{ij} (\g^\m)_{ab} \pa_\m B ~~, \\ 
\left\{\text{D}^{i}_{a},\hat{\text{D}}^{j}_{b}\right\} \breve{B} =& - 2 i (\s^2)^{ij} (\g^5\g^\m)_{ab} \pa_\m 
A - 2 i (\s^2)^{ij} (\g^5)_{ab} F + \fracm 12 \delta^{ij} (\g^5 [\g^\m, \g^\n])_{ab} F_{\m\n} ~~, \\ 
\left\{\text{D}^{i}_{a},\hat{\text{D}}^{j}_{b}\right\}  \hat{F} = & 2 i (\s^3)^{ij} (\g^\m)_{ab} \pa_\m F + 2 i 
(\s^2)^{ij} (\g^5 \g^\m)_{ab} \pa_\m \rd - 2 i (\s^1)^{ij} (\g^\m)_{ab} \pa^\n F_{\m\n}~~, \\ 
\left\{\text{D}^{i}_{a},\hat{\text{D}}^{j}_{b}\right\}  \breve{F} = & 2 i (\s^1)^{ij} (\g^\m)_{ab} \pa_\m F - 2 i 
(\s^2)^{ij} (\g^5 \g^\m)_{ab} \pa_\m G + 2 i (\s^3)^{ij} (\g^\m)_{ab} \pa^\n F_{\m\n} ~~, \\ 
\left\{\text{D}^{i}_{a},\hat{\text{D}}^{j}_{b}\right\} \hat{G} = & 2 i (\s^3)^{ij} (\g^\m)_{ab} \pa_\m G + 2 i 
(\s^1)^{ij} (\g^\m)_{ab} \pa_\m \rd + 2 i (\s^2)^{ij}(\g^5\g^\m)_{ab} \pa^\n F_{\m\n} ~~, \\ 
\left\{\text{D}^{i}_{a},\hat{\text{D}}^{j}_{b}\right\} \breve{G} = & - 2 i (\s^2)^{ij} (\g^5)_{ab} \square A + 
2 (\s^2)^{ij} C_{ab} \square B - 2 i (\s^2)^{ij} (\g^5\g^\m)_{ab} \pa_\m F \cr
&- 2 i (\s^1)^{ij} (\g^\m)_{ab} \pa_\m G + 2 i (\s^3)^{ij} (\g^\m)_{ab} \pa_\m \rd  ~~~.
\end{align}The gauge term in the above for the vector field $A_\m$ is
\begin{align}
	\hat{\lambda}^{ij}_{ab} = 2 i (\s^2)^{ij} (\g^5)_{ab} \hat{B}~~~.
\end{align}
For the fermions we have
\begin{align}
 \{\text{D}^{i}_{a},\hat{\text{D}}^{j}_{b}\} \Psi^k_c ~&= 2 i \hat{Z}_1^{ijkl} (\g^\mu)_{ab} \pa_\mu \hat{\Psi}^l_c 
 + 2 i \hat{Z}_2^{ijkl} ([\g^\mu, \g^\nu])_{ab}(\g_\nu)_c{}^d \pa_\mu \hat{\Psi}^l_d \cr
 &{~~~} + 2 i \hat{Z}_3^{ijkl} (\g_\nu )_{ab}([\g^\nu, \g^\mu])_c{}^d \pa_\mu \hat{\Psi}^l_d +2 i \hat{Z}_4^{ijkl} (\g^5
 [\g^\mu, \g^\nu])_{ab}(\g^5\g_\nu)_c{}^d \pa_\mu \hat{\Psi}^l_d \cr
 &{~~~} +2 i \hat{Z}_5^{ijkl} (\g^5 \g^\mu)_{ab} (\g^5)_c{}^d \pa_\mu \hat{\Psi}^l_d + 2 i \hat{Z}_6^{ijkl} (\g^5\g_\nu 
 )_{ab}(\g^5[\g^\nu, \g^\mu])_c{}^d \pa_\mu \hat{\Psi}^l_d \cr
 &{~~~} + 2 i \hat{Z}_7^{ijkl} (\g^5)_{ab} (\g^5 \g^\mu)_c{}^d \pa_\mu \hat{\Psi}_d^l + 2 i \hat{Z}_8^{ijkl} C_{ab} 
 (\g^\mu)_c{}^d \pa_\mu \hat{\Psi}_d^l  ~~, \\ 
 \{\text{D}^{i}_{a},\hat{\text{D}}^{j}_{b}\} \hat{\Psi}^k_c ~&= 2 i \hat{\hat{Z}}_1^{ijkl} (\g^\mu)_{ab} \pa_\mu 
 \Psi^l_c + 2 i \hat{\hat{Z}}_2^{ijkl} ([\g^\mu, \g^\nu])_{ab}(\g_\nu)_c{}^d \pa_\mu \Psi^l_d \cr
 &{~~~} + 2 i \hat{\hat{Z}}_3^{ijkl} (\g_\nu )_{ab}([\g^\nu, \g^\mu])_c{}^d \pa_\mu \Psi^l_d +2 i \hat{\hat{Z}}_4^{ijkl} 
 (\g^5[\g^\mu, \g^\nu])_{ab}(\g^5\g_\nu)_c{}^d \pa_\mu \Psi^l_d \cr
 &{~~~} +2 i \hat{\hat{Z}}_5^{ijkl} (\g^5 \g^\mu)_{ab} (\g^5)_c{}^d \pa_\mu \Psi^l_d + 2 i \hat{\hat{Z}}_6^{ijkl} (\g^5
 \g_\nu )_{ab}(\g^5[\g^\nu, \g^\mu])_c{}^d \pa_\mu \Psi^l_d \cr
 &{~~~} + 2 i \hat{\hat{Z}}_7^{ijkl} (\g^5)_{ab} (\g^5 \g^\mu)_c{}^d \pa_\mu \Psi_d^l + 2 i \hat{\hat{Z}}_8^{ijkl} 
 C_{ab} (\g^\mu)_c{}^d \pa_\mu \Psi_d^l   ~~~,
\end{align}
where the ``$Z$-factors'' take the forms,
\begin{align}
\begin{split}
\hat{Z}_1^{\text{ijkl}}=&\frac{3}{4} \left(\s ^{1}\right)^{\text{ik}} \left(\s ^{1}\right)^{\text{jl}}+\frac{1}{4} \left(\s^{1}
\right)^{\text{il}} \left(\s ^{1}\right)^{\text{jk}}+\left(\s ^3\right)^{\text{il}} \left(\s^{3}\right)^{\text{jk}}~~, \cr
\hat{Z}_2^{\text{ijkl}}=&\frac{1}{8} \left(\s ^{1}\right)^{\text{il}} \left(\s^{1}\right)^{\text{jk}}-\frac{1}{8} 
\left(\s^{1}\right)^{\text{ik}} \left(\s^{1}\right)^{\text{jl}}~~, \cr
\hat{Z}_3^{\text{ijkl}}=&\frac{1}{8} \left(\s ^{1}\right)^{\text{ik}} \left(\s^{1}\right)^{\text{jl}}-\frac{1}{8} 
\left(\s ^{1}\right)^{\text{il}} \left(\s^{1}\right)^{\text{jk}}~~, \cr
\hat{Z}_4^{\text{ijkl}}=&\frac{3}{8} \left(\s ^{1}\right)^{\text{il}} \left(\s^{1}\right)^{\text{jk}}-\frac{3}{8} 
\left(\s ^{1}\right)^{\text{ik}} \left(\s^{1}\right)^{\text{jl}}~~, \cr
\hat{Z}_5^{\text{ijkl}}=&-\frac{1}{2} \left(\s ^{1}\right)^{\text{ij}} \left(\s^{1}\right)^{\text{kl}}+\frac{1}{4} 
\left(\s ^{1}\right)^{\text{ik}} \left(\s ^{1}\right)^{\text{jl}}+\frac{1}{4} \left(\s ^{1}\right)^{\text{il}} \left(\s ^{1}
\right)^{\text{jk}}~~, \cr
\hat{Z}_6^{\text{ijkl}}=&\frac{1}{4} \left(\s^{1}\right)^{\text{ij}} \left(\s ^{1}\right)^{\text{kl}}-\frac{1}{8} 
\left(\s ^{1}\right)^{\text{ik}} \left(\s^{1}\right)^{\text{jl}}-\frac{1}{8} \left(\s ^{1}\right)^{\text{il}} \left(\s^{1}
\right)^{\text{jk}}~~, \cr
\hat{Z}_7^{\text{ijkl}}=&-\frac{1}{2} \left(\s ^{1}\right)^{\text{ij}} \left(\s^{1}\right)^{\text{kl}}+\frac{1}{4} 
\left(\s ^{1}\right)^{\text{ik}} \left(\s ^{1}\right)^{\text{jl}}+\frac{1}{4} \left(\s ^{1}\right)^{\text{il}} \left(\s^{1}
\right)^{\text{jk}}~~, \cr
\hat{Z}_8^{\text{ijkl}}=&-\frac{1}{2} \left(\s^{1}\right)^{\text{ij}} \left(\s ^{1}\right)^{\text{kl}}-\frac{1}{4} 
\left(\s^{1}\right)^{\text{ik}} \left(\s^{1}\right)^{\text{jl}}+\frac{3}{4} \left(\s ^{1}\right)^{\text{il}} \left(\s^{1}
\right)^{\text{jk}}
\end{split}
\end{align}

\begin{align}
\begin{split}
\hat{\hat{Z}}_1^{\text{ijkl}}=&\frac{5}{4} \left(\s ^1\right)^{\text{ik}} \left(\s ^1\right)^{\text{jl}}-\frac{1}{4} 
\left(\s^1\right)^{\text{il}} \left(\s ^1\right)^{\text{jk}}+\left(\s ^3\right)^{\text{il}} \left(\s^3\right)^{\text{jk}}
~~,   {~~~~~~~} \cr
\hat{\hat{Z}}_2^{\text{ijkl}}=&\frac{1}{8} \left(\s ^1\right)^{\text{ik}} \left(\s^1\right)^{\text{jl}}-\frac{1}{8} 
\left(\s^1\right)^{\text{il}} \left(\s^1\right)^{\text{jk}}~~, \cr
\hat{\hat{Z}}_3^{\text{ijkl}}=&\frac{1}{8} \left(\s ^1\right)^{\text{il}} \left(\s^1\right)^{\text{jk}}-\frac{1}{8} 
\left(\s^1\right)^{\text{ik}} \left(\s^1\right)^{\text{jl}}~~, \cr
\hat{\hat{Z}}_4^{\text{ijkl}}=&\frac{3}{8} \left(\s ^1\right)^{\text{ik}} \left(\s^1\right)^{\text{jl}}-\frac{3}{8} 
\left(\s^1\right)^{\text{il}} \left(\s^1\right)^{\text{jk}}~~, \cr
\end{split}
\end{align}

\begin{align}
\begin{split}
\hat{\hat{Z}}_5^{\text{ijkl}}=&\frac{1}{2} \left(\s ^1\right)^{\text{ij}} \left(\s^1\right)^{\text{kl}}-\frac{1}{4} 
\left(\s^1\right)^{\text{ik}} \left(\s ^1\right)^{\text{jl}}-\frac{1}{4} \left(\s ^1\right)^{\text{il}} \left(\s ^1
\right)^{\text{jk}}~~, \cr
\hat{\hat{Z}}_6^{\text{ijkl}}=&-\frac{1}{4} \left(\s ^1\right)^{\text{ij}} \left(\s ^1\right)^{\text{kl}}+
\frac{1}{8} \left(\s ^1\right)^{\text{ik}} \left(\s ^1\right)^{\text{jl}}+\frac{1}{8} \left(\s ^1\right)^{\text{il}} 
\left(\s ^1\right)^{\text{jk}}~~, \cr
\hat{\hat{Z}}_7^{\text{ijkl}}=&\frac{1}{2} \left(\s^1\right)^{\text{ij}} \left(\s ^1\right)^{\text{kl}}-
\frac{1}{4} \left(\s^1\right)^{\text{ik}} \left(\s^1\right)^{\text{jl}}-\frac{1}{4} \left(\s ^1\right)^{\text{il}} 
\left(\s^1\right)^{\text{jk}}~~, \cr
\hat{\hat{Z}}_8^{\text{ijkl}}=&\frac{1}{2} \left(\s ^1\right)^{\text{ij}} \left(\s^1\right)^{\text{kl}}-
\frac{3}{4} \left(\s^1\right)^{\text{ik}} \left(\s ^1\right)^{\text{jl}}+\frac{1}{4} \left(\s^1\right)^{\text{il}} 
\left(\s ^1\right)^{\text{jk}} ~~~.
\end{split}
\end{align}

To compare the results seen here against the ones found in chapter two for
the 4D, $\cal N$ = 4 vector-tensor supermultiplet, the form of the super algebra 
of its four supercharges ${{\text D}}{}^{i}_{a}$ and ${\hat {\text D}}{}^{i}_{a}$
in this chapter is
\be \eqalign{
 \{{\text{D}}^{i}_{a},{\text{D}}^{j}_{b}\}  ~&=~ i 2 \delta^{ij} (\g^{\mu})_{ab}  \pa_{\mu}\,+ \,    
 {\rm {central ~charge}}   
 ~~~, \cr  
\{  \hat{\text{D}}^{i}_{a} , \hat{\text{D}}^{j}_{b}  \}  ~&=~i 2 \delta^{ij} (\g^{\mu})_{ab}  
\pa_{\mu} \,+ \,   {\rm {central ~charge}} ~~~, \cr
\{{\text{D}}^{i}_{a}, \hat{\text{D}}^{j}_{b}  \}  ~&=~ {\rm {central ~charge}} ~~~,
}\ee
on the fields in the W-F 4D, $\cal N$ = 2 submultiplet.  An examination of the 
results in (\ref{e:CChathatAlgebra}) and in (\ref{e:CCAlgebra}) reveals that the central charges that appear in 
the first two equations above are identical by our choice of the ansatz in (\ref{e:Wansatz})
and (\ref{e:Xansatz}).  The four supercharges close on the fields of the 4D, $\cal N$ = 2 
vector submultiplet without central charges.

\section{Conclusion}
\label{conclusions}

 \vskip,2in

This short note expressed the 4D, $\cal N$=4 vector-tensor supermultiplet in terms of a GL(2,$\mathbb{R}$)$\otimes$GL(2,$\mathbb{R}$) isospin structure, rather than the original SP(4) isospin structure presented in~\cite{Sohnius:1980it}. The GL(2,$\mathbb{R}$)$\otimes$GL(2,$\mathbb{R}$) structure maintains the manifest off-shell closure of the 4D, $\cal N$ = 2 vector and tensor submultiplets. Using modern computing techniques, we have exhaustively analyzed all possible ways of marrying the two 4D, $\cal N$ = 2 off-shell supermultiplets discussed in this paper into a 4D, $\cal N$=4 off-shell multiplets, and found no possible way of closing the resulting $\cal N$=4 algebra without central charges. The associated \emph{Mathematica} code is available open-source at the \href{https://hepthools.github.io/Data/}{HEPTHools Data Repository}.

Thus, our work has extended a familiar construction that has long been used in
supersymmetrical field theories.  One can begin with two 4D, $\cal N$ = 1
chiral supermultiplets in the context of a non-linear $\s$-model and extend
this to an $\cal N$ = 2 non-linear $\s$-model by introducing a complex
structure $f{}_j{}^j$.  In the current discussion, the analogs of such a complex
structure consist of the set of twelve matrices shown in~(\ref{e:V}) for combining
the 4D, $\cal N$ = 2 vector and tensor supermultiplets to form a 4D, $\cal N$ 
= 4 representation or the set of twelve matrices shown in (\ref{e:X}) for combining
the 4D, $\cal N$ = 2 vector and W-F supermultiplets to form a 4D, $\cal N$ 
= 4 representation.  So this immediately raises the question of precisely
what mathematical structure is being described by these set of matrices?

There are several possible future avenues that suggest themselves for
further study.  One obvious one is the mathematical structure of non-linear
$\s$-models related to these 4D, $\cal N$ = 4 supermultiplets.  This is a
question to be answered both in four dimensions and in the dimensional
reduction of such models.  Another distinct direction is to use these
sorts of discussions to drive exploration of the representation theory
of supersymmetrical model via the approach of adinkras \cite{GRana1,GRana2,adnk1,Gates:2009me,Gates:2011aa,Gates:2012zr,Chappell:2012qf}
and corresponding methods in four dimensional theories\cite{Gates:2012xb,Gates:2014npa,Calkins:2014sma,Holor4D1,Gates:2016bzr}.  

It should be clear that the doublets of supercovariant derivatives either 
given by (${\text{D}}^{i}_{a}, {\tilde{\text{D}}}^{j}_{b}$) or (${\text{D}}^{i}_{a}, 
\hat{\text{D}}^{j}_{b}$) can be regarded as the components of a single
supercovariant derivative ${\text{D}}^{i \, {\Hat A}}_{a}$ that possesses
a pair of ``isospin'' indices that each take on two values.  Given that the 
4D, $\cal N$ = 4 vector-tensor supermultiplet possesses the simplest 
central charge structure, it might be profitable to study the supervector 
fields associated with the supermanifold whose coordinates are dual to 
${\text{D}}^{i \, {\Hat A}}_{a}$ in the context of the 4D, $\cal N$ = 4 
superconformal symmetry.

 \vspace{.05in}
 \begin{center}
\parbox{4in}{{\it ``If you obey all the rules, you miss all the fun.'' \\ ${~}$ 
\\ ${~}$ }\,\,-\,\, Katharine Hepburn $~~~~~~~~~$}
 \parbox{4in}{
 $~~$}  
 \end{center}

 \noindent
{\bf Acknowledgments}\\[.1in] \indent
 This research is supported by the 
endowment of the Ford Foundation Professorship of Physics at 
Brown University and partially supported by the U.S. National 
Science Foundation grant PHY-1315155.

\appendix
\section{Reviewing 4D, \texorpdfstring{$\mathcal{N}$}{N} = 2 Supersymmetry Results}\label{a:N2Review}

In this appendix, for the convenience of the reader we simply present the form of
the 4D, $\cal N$ = 2 supermultiplets used in our text.  We use the index convention
$i = 1,2$ labels the two supersymmetries.  Furthermore our definitions are such
that
\begin{align}
   (\s^0)^{ij}=\delta^{ij} &=\left(\begin{array}{l l}
                              1 & 0 \\
                              0 & 1 
                   \end{array}
              \right) ~~,~~
              (\s^1)^{ij} = \left(\begin{array}{l l}
                              0 & 1 \\
                              1 & 0 
                   \end{array} \right)
              ~~~,~~~
              (\s^2)^{ij} = \left(\begin{array}{l l}
                              0 & -i \\
                              i & 0 
                   \end{array}
              \right)~~~,~~~
              (\s^3)^{ij} = \left(\begin{array}{l l}
                              1 & 0 \\
                              0 & -1 
                   \end{array}
              \right)~~~, \nonumber\\
& {~~~~~~~~~~~~~~~~~~} \eta_{\mu\nu} = \left(\begin{array}{l l l l}
                           -1 & 0 & 0 & 0\\
                            0 & 1 & 0 & 0\\
                            0 & 0 & 1 & 0\\
                            0 & 0 & 0 & 1
                        \end{array}
                 \right)  ~~~,~~~
\epsilon_{0123} = 1 ~~~ .
\end{align}

\subsection{4D, \texorpdfstring{$\mathcal{N}$}{N}=2 Vector Supermultiplet}

The corresponding transformation laws are
\be
\begin{split}
 \text{D}^{i}_{a} A ~&=~ \Psi^{i}_{a}  ~~~, ~~~
 \text{D}^{i}_{a} B ~=~ i (\g^{5})_{a}{}^{b} \Psi^{i}_{b}  ~~~, ~~~
 \text{D}^{i}_{a} F ~=~ (\g^{\mu})_{a}{}^{b} \pa_{\mu} \Psi^{i}_{b}  ~~~, \\
 \text{D}^{i}_{a} G ~&=~ i (\s^{3})^{ij} (\g^{5}\g^{\mu})_{a}{}^{b} \pa_{\mu} \Psi^{j}_{b}  ~~~, ~~~
 \text{D}^{i}_{a} A_{\mu} ~=~ i (\s^{2})^{ij} (\g_{\mu})_{a}{}^{b} \Psi^{j}_{b}  ~~~, ~~~
 \text{D}^{i}_{a} d ~=~ i (\s^{1})^{ij} (\g^{5}\g^{\mu})_{a}{}^{b} \pa_{\mu} \Psi^{j}_{b}  ~~~, \\
 \text{D}^{i}_{a} \Psi^{j}_{b} ~&=~ \delta^{ij} \big[ i (\g^{\mu})_{ab} \pa_{\mu} A - (\g^{5}\g^{\mu})_{ab} \pa_{\mu} B - i C_{ab} F \big] + (\s^{3})^{ij} (\g^{5})_{ab} G  ~~~, \\
 & {~~~~~}+ (\s^{1})^{ij} (\g^{5})_{ab} d + \tfrac{1}{4} (\s^{2})^{ij} ([\g^{\mu},\g^{\nu}])_{ab} (\pa_{\mu} A_{\nu} - \pa_{\nu} A_{\mu})  ~~~.
\end{split}
\ee
The transformation laws satisfy the algebra
\begin{align}\label{e:N2VectorClose}
 \{\text{D}^{i}_{a},\text{D}^{j}_{b}\} \chi ~&=~ i 2 \delta^{ij} (\g^{\mu})_{ab} \pa_{\mu} \chi  ~~~,  \\ 
 \label{e:N2VectorCloseGauge}
 \{\text{D}^{i}_{a},\text{D}^{j}_{b}\} A_{\mu} ~&=~ i 2 \delta^{ij} (\g^{\nu})_{ab} F_{\nu\mu} + i (\s^{2})^{ij} \left[ i 2 C_{ab} \pa_{\mu} A - 2(\g^{5})_{ab} \pa_{\mu} B  \right]   ~~~,
\end{align}
where
$
 \chi \in \{A,B,F,G,d,\Psi^{k}_{c}\}
$
and the Lagrangian is 
\be
\begin{split}
 \mathcal{L}{}_{(2VS)} =& - \tfrac{1}{2} \pa_{\mu} A \pa^{\mu} A - \tfrac{1}{2} \pa_{\mu} B \pa^{\mu} B + \tfrac{1}{2} F^{2} + \tfrac{1}{2} G^{2} - \tfrac{1}{4} F_{\mu\nu} F^{\mu\nu} + \tfrac{1}{2} d^{2} 
 \\&  + i \tfrac{1}{2} (\g^{\mu})^{bc} \Psi^{i}_{b} \pa_{\mu} \Psi^{i}_{c}   ~~~, 
\end{split}
\ee
with
$
F_{\mu\nu} = \pa_{\mu} A_{\nu} - \pa_{\nu} A_{\mu}
$.

\subsection{4D, \texorpdfstring{$\mathcal{N}$}{N}=2 Tensor Supermultiplet}

The corresponding transformation laws are
\be
\begin{split}
 \text{D}^{i}_{a} {\Tilde A} ~&=~ (\s^{3})^{ij} {\Tilde {\Psi}}^{j}_{a}  ~~~,~~~
 \text{D}^{i}_{a} {\Tilde B} ~=~ i (\g^{5})_{a}{}^{b} {\Tilde {\Psi}}^{i}_{b}  ~~~, ~~~
 \text{D}^{i}_{a} {\Tilde F} ~=~ (\g^{\mu})_{a}{}^{b} \pa_{\mu} {\Tilde {\Psi}}^{i}_{b}  ~~~, \\
 \text{D}^{i}_{a} {\Tilde G} ~&=~ i  (\g^{5}\g^{\mu})_{a}{}^{b} \pa_{\mu} {\Tilde {\Psi}}^{i}_{b}  ~~~, ~~~
 \text{D}^{i}_{a} {\Tilde {\varphi}} ~=~ (\s^{1})^{ij} {\Tilde {\Psi}}^{j}_{a}  ~~~, ~~~
 \text{D}^{i}_{a} {\Tilde B}_{\mu\nu} ~=~ - i \tfrac{1}{4} (\s^{2})^{ij} ([\g_{\mu},\g_{\nu}])_{a}{}^{b} {\Tilde {\Psi}}^{j}_{b}  ~~~, \\
 \text{D}^{i}_{a} {\Tilde {\Psi}}^{j}_{b} ~&=~ \delta^{ij} \big[ - (\g^{5}\g^{\mu})_{ab} \pa_{\mu} {\Tilde B} - i C_{ab} {\Tilde F} + (\g^{5})_{ab} {\Tilde G}  \big] + i (\s^{3})^{ij} (\g^{\mu})_{ab} \pa_{\mu} {\Tilde A}  ~~~, \\
 &{~~~}  + i (\s^{1})^{ij} (\g^{\mu})_{ab} \pa_{\mu} {\Tilde {\varphi}} - i (\s^{2})^{ij} \epsilon_{\mu}{}^{\nu\alpha\beta} (\g^{5}\g^{\mu})_{ab} \pa_{\nu} {\Tilde B}_{\alpha\beta}  ~~~,
\end{split} \ee
The transformation laws satisfy the algebra
\begin{align}\label{e:N2TensorClose}
 \{\text{D}^{i}_{a},\text{D}^{j}_{b}\} \chi ~=&~ i 2 \delta^{ij} (\g^{\mu})_{ab} \pa_{\mu} \chi  ~~~,  \\
 \label{e:N2TensorCloseGauge}
\begin{split}
 \{\text{D}^{i}_{a},\text{D}^{j}_{b}\} {\Tilde B}_{\mu\nu} ~=&~ i 2 \delta^{ij} (\g^{\alpha})_{ab} H_{\alpha\mu\nu} 
 \\ &  + i (\g_{[\mu})_{ac}\pa_{\nu]} \left[ (\s^{1})^{ij} \delta_{b}{}^{c} {\Tilde A} + (\s^{2})^{ij} (\g^{5})_{b}{}^{c} {\Tilde B} - (\s^{3})^{ij} \delta_{b}{}^{c} {\Tilde {\varphi}}  \right]    ~~~, 
\end{split}
\end{align}
where
$
 \chi \in \{{\Tilde A},{\Tilde B},{\Tilde F},{\Tilde G},{\Tilde {\varphi}},{\Tilde {\Psi}}^{k}_{c}\} 
$
and the Lagrangian is 
\be
\begin{split}
 \mathcal{L}{}_{(2TS)} =& - \tfrac{1}{2} \pa_{\mu} {\Tilde A} \pa^{\mu} {\Tilde A} - \tfrac{1}{2} \pa_{\mu} {\Tilde B} \pa^{\mu} {\Tilde B} + \tfrac{1}{2} {\Tilde F}^{2} + \tfrac{1}{2} {\Tilde G}^{2} - \tfrac{1}{3} H_{\mu\nu\alpha} H^{\mu\nu\alpha} - \tfrac{1}{2} \pa_{\mu} {\Tilde {\varphi}} \pa^{\mu} {\Tilde {\varphi}}
 \\&  + i \tfrac{1}{2} (\g^{\mu})^{bc} {\Tilde \Psi}^{i}_{b} \pa_{\mu} {\Tilde \Psi}^{i}_{c}   ~~~,
\end{split}
\ee
with
$
 H_{\mu\nu\alpha} = \pa_{\mu} {\Tilde B}_{\nu\alpha} + \pa_{\nu} {\Tilde B}_{\alpha\mu} + \pa_{\alpha} {\Tilde B}_{\mu\nu}
$.

\subsection{4D, \texorpdfstring{$\mathcal{N}$}{N}=2 W-F Hypermultiplet}
The transformation laws for the Wess-Fayet (W-F) hypermultiplet containing the Chiral-Chiral 
multiplet combination are
\begin{align}
\begin{split}
  {\rm D}_a^i \hat{A} &= (\s^3)^{ij}\hat{\Psi}_a^j, ~~~, \\
  {\rm D}_a^i \hat{B} &= i (\g^5)_a^{~b} \hat{\Psi}_b^i, ~~~, \\
  {\rm D}_a^i \hat{F} &= (\s^3)^{ij}(\g^{\mu})_a^{~b} \pa_{\mu}\hat{\Psi}_b^j,~~~, \\
  {\rm D}_a^i \hat{G} &= i (\g^5\g^\mu)_a^{~b}\pa_\mu \hat{\Psi}_b^i,~~~, \\
  {\rm D}_a^i \breve{A} &= (\s^1)^{ij}\hat{\Psi}_a^j, ~~~, \\
  {\rm D}_a^i \breve{B} &= -(\s^2)^{ij} (\g^5)_a^{~b} \hat{\Psi}_b^j,~~~, \\
  {\rm D}_a^i \breve{F} &= (\s^1)^{ij}(\g^{\mu})_a^{~b} \pa_{\mu}\hat{\Psi}_b^j,~~~, \\
  {\rm D}_a^i \breve{G} &= -(\s^2)^{ij} (\g^5\g^\mu)_a^{~b}\pa_\mu \hat{\Psi}_b^j,~~~, \\
  {\rm D}_a^i \hat{\Psi}_b^j &= i(\s^3)^{ij}\left((\g^{\mu})_{ab}\pa_{\mu}\hat{A}-C_{ab}\hat{F}\right) + 
  \delta^{ij}(-(\g^5\g^{\mu})_{ab}\pa_{\mu}\hat{B} + (\g^5_{ab}) \hat{G}) ~~~, \\
  &{~~~} +i(\s^1)^{ij}\left((\g^{\mu})_{ab}\pa_{\mu}\breve{A}-C_{ab}\breve{F}\right) +
  i(\s^2)^{ij}(-(\g^5\g^{\mu})_{ab}\pa_{\mu}\breve{B} + (\g^5_{ab})\breve{G}) ~~~,
 \end{split}
\end{align}
The following Lagrangian is invariant with respect to these transformations:
\begin{align}
\mathcal{L}_{(2CC)} &= -\frac{1}{2} \pa_{\mu}\hat{A} \pa^\mu \hat{A} -\frac{1}{2} \pa_{\mu}\breve{A} 
\pa^\mu \breve{A}-\frac{1}{2} \pa_{\mu}\hat{B} \pa^\mu \hat{B}-\frac{1}{2} \pa_{\mu}\breve{B} 
\pa^\mu \breve{B} + \nonumber\\
&~+\frac{1}{2} \hat{F}^2 + \frac{1}{2} \breve{F}^2 + \frac{1}{2} \hat{G}^2 + \frac{1}{2} \breve{G}^2 +\frac{1}{2}i
(\g^\mu)^{cd}\hat{\Psi}^i_c \pa_\mu \hat{\Psi}^i_d  ~~~,
\end{align}
which easily seen to be the direct sum of the $\cal N$ $=$ 1 invariant Lagrangians for
the separate ($\hat{A}, \hat{B} \, \hat{\Psi}^1_c , \, \hat{F}, \hat{G} $) chiral supermultiplet and the ($\breve{A}, \breve{B}, \, \hat{\Psi}^2_c , \, \breve{F}, \breve{G}$) chiral supermultiplet.  Direct calculation yields the following algebra:
\begin{align}\label{e:CCAlgebra}
\begin{split}
\left\{{\rm D}^i_a,{\rm D}^j_b\right\}\hat{A}=&\delta^{ij}2i\left(\g^\mu\right)_{ab}\pa_\mu \hat{A}+i\left(\s^2\right)^{ij}2iC_{ab}\breve{F}~~~, \\
\left\{{\rm D}^i_a,{\rm D}^j_b\right\}\breve{A}=&\delta^{ij}2i\left(\g^\mu\right)_{ab}\pa_\mu\breve{A}-i\left(\s^2\right)^{ij}2iC_{ab}\hat{F}~~~, \\
\left\{{\rm D}^i_a,{\rm D}^j_b\right\}\hat{B}=&\delta^{ij}2i\left(\g^\mu\right)_{ab}\pa_\mu \hat{B}+i\left(\s^2\right)^{ij}2iC_{ab}\breve{G}~~~, \\
\left\{{\rm D}^i_a,{\rm D}^j_b\right\}\breve{B}=&\delta^{ij}2i\left(\g^\mu\right)_{ab}\pa_\mu\breve{B}-i\left(\s^2\right)^{ij}2iC_{ab} \hat{G}~~~, \\
\left\{{\rm D}^i_a,{\rm D}^j_b\right\}\hat{F}=&\delta^{ij}2i\left(\g^\mu\right)_{ab}\pa_\mu \hat{F}+i\left(\s^2\right)^{ij}2iC_{ab}\square\breve{A}~~~, \\
\left\{{\rm D}^i_a,{\rm D}^j_b\right\}\breve{F}=&\delta^{ij}2i\left(\g^\mu\right)_{ab}\pa_\mu\breve{F}-i\left(\s^2\right)^{ij}2iC_{ab}\square \hat{A}~~~, \\
\left\{{\rm D}^i_a,{\rm D}^j_b\right\}\hat{G}=&\delta^{ij}2i\left(\g^\mu\right)_{ab}\pa_\mu \hat{G}+i\left(\s^2\right)^{ij}2iC_{ab}\square\breve{B}~~~, \\
\left\{{\rm D}^i_a,{\rm D}^j_b\right\}\breve{G}=&\delta^{ij}2i\left(\g^\mu\right)_{ab}\pa_\mu\breve{G}-i\left(\s^2\right)^{ij}2iC_{ab}\square \hat{B}~~~, \\
\left\{{\rm D}^i_a,{\rm D}^j_b\right\}\hat{\Psi}^k_c=&\delta^{ij}2i\left(\g^\mu\right)_{ab}\pa_\mu \hat{\Psi}^k_c-\left(\s^2\right)^{ij}\left(\s^2\right)^{kr}2iC_{ab}\left(\g^\mu\right)_{c}^{~d}\pa_\mu\hat{\Psi}^r_d  ~~~.
\end{split}
\end{align}
As can easily be seen (and is well-known) the pair-results in (\ref{e:N2VectorClose})-(\ref{e:N2VectorCloseGauge}) and as well as the pair-results
(\ref{e:N2TensorClose})-(\ref{e:N2TensorCloseGauge}) describe closure of the SUSY algebra {\em {without}} central charges nor use of equations
of motion for fermion fields.  This is very different than the results in (\ref{e:CCAlgebra}) where the closure of the
algebra requires both the presence of central charges realized on the bosons and the enforcement
of equations of motion on the fermions.

\section{Reviewing 4D, \texorpdfstring{$\mathcal{N}$}{N} = 4 Supersymmetry Results}\label{a:N4Review}
In this section, we review the Abelian 4D, $\mathcal{N}=4$ SUSY-YM system in a Majarana representation as presented in~\cite{Gates:2015dda}.

\subsection{\texorpdfstring{${\mathcal N} = 4$}{{\mathcal N} = 4} Transformation Laws}\label{CCCVReductionNequals4TLaws}
The Lagrangian for the Abelian $d=4$, ${\mathcal N} = 4$ SUSY-YM system
\be\label{eq:SUSYYMLagrangian}\eqalign{
{\mathcal{L}} = &-\frac{1}{2}(\partial_{\mu}A^{J})(\partial^{\mu}A^{J}) -\frac{1}{2}(\partial_{\mu}B^{J})(\partial^{\mu}B^{J})\cr
&+i\frac{1}{2}(\gamma^{\mu})^{ab}\psi_{a}^{J}\partial_{\mu}\psi_{b}^{J} +\frac{1}{2}(F^{J})^{2}+\frac{1}{2}(G^{J})^{2}\cr
&-\frac{1}{4}F_{\mu\nu}F^{\mu\nu} +\frac{1}{2}i(\gamma^{\mu})^{cd}\lambda_{c}\partial_{\mu}\lambda_{d}+\frac{1}{2}{\rm d}^2
}\ee
is invariant with respect to the global supersymmetric transformations
\be\eqalign{
{\rm D}_a A^J ~=~&~ \psi_a^J  ~~~, \cr
{\rm D}_a B^J ~=~ &~i \, (\gamma^5){}_a{}^b \, \psi_b^J  ~~~, \cr
{\rm D}_a \psi_b^J ~=~ &~i\, (\gamma^\mu){}_{a \,b}\,  \partial_\mu A^J 
~-~ ~ (\gamma^5\gamma^\mu){}_{a \,b} \, \partial_\mu B^J ~\cr
&-~ ~i \, C_{a\, b} 
\,F^J  ~+~ ~ (\gamma^5){}_{ a \, b} G^J  ~~, \cr
{\rm D}_a F^J ~=~&~ (\gamma^\mu){}_a{}^b \, \partial_\mu \, \psi_b^J   ~~~, \cr
{\rm D}_a G^J ~=~ &~i \,(\gamma^5\gamma^\mu){}_a{}^b \, \partial_\mu \,  
\psi_b^J  ~~~.
} \label{eq:chi0specific}
\ee
\\
\be \eqalign{
{\rm D}_a \, A{}_{\mu} ~=~ & (\gamma_\mu){}_a {}^b \,  \l_b  ~~~, \cr
{\rm D}_a \l_b ~=~ &  - \, \, \fracm 12 (\sigma^{\mu\nu})_a{}_b \, F_{\mu\nu}
~+~  (\gamma^5){}_{a \,b} \,    {\rm d} ~~,  \cr
{\rm D}_a \, {\rm d} ~=~ & i \, (\gamma^5\gamma^\mu){}_a {}^b \, 
\,  \partial_\mu \l_b  ~~~. \cr
} \label{eq:V0specific}
\ee
\\
\be \eqalign{
{\rm D}_a^{I} A^J ~=~& \delta^{IJ}~ \l_a - \epsilon^{IJ}_{~~K} \psi_a^K~~~, \cr
{\rm D}_a^{I} B^J ~=~& ~i \, (\gamma^5){}_a{}^b \, [~\delta^{IJ}~ \l_b +\epsilon^{IJ}_{~~K} \psi_b^K~]  ~~~, \cr
{\rm D}_a^{I} \psi_b^J ~=~& \delta^{IJ}[~\, \, \fracm 12 (\sigma^{\mu\nu})_{ab} \,F_{\mu\nu}
~+~  (\gamma^5){}_{a \,b} \,    {\rm d}~]\cr
&-~\epsilon^{IJ}_{~~K}[~
~-i\, (\gamma^\mu){}_{a \,b}\,  \partial_\mu A^K 
~-~ (\gamma^5\gamma^\mu){}_{a \,b} \, \partial_\mu B^K ~\cr
&+~ i \, C_{a\, b} 
\,F^K  ~+~ (\gamma^5){}_{ a \, b} G^K~]  ~~, \cr
{\rm D}_a^{I} F^J ~=~& ~ (\gamma^\mu){}_a{}^b \, \partial_\mu \, [~\delta^{IJ}~ \l_b - \epsilon^{IJ}_{~~K} \psi_b^K~]   ~~~, \cr
{\rm D}_a^{I} G^J ~=~& ~i \,(\gamma^5\gamma^\mu){}_a{}^b \, \partial_\mu \,  
[~-\delta^{IJ}~ \l_b + \epsilon^{IJ}_{~~K} \psi_b^K~]  ~~~.
}\label{eq:chiIspecific}
\ee
\\
\be \eqalign{
{\rm D}_a^{I} \, A{}_{\mu} ~=~ & -(\gamma_\mu){}_a {}^b \,  \psi_b^{I}  ~~~, \cr
{\rm D}_a^{I} \l_b ~=~ & ~i\, (\gamma^\mu){}_{a \,b}\,  \partial_\mu A^I 
~-~ ~ (\gamma^5\gamma^\mu){}_{a \,b} \, \partial_\mu B^I ~\cr
&-~i \, C_{a\, b} 
\,F^I  ~-~ (\gamma^5){}_{ a \, b} G^I  ~~, \cr
{\rm D}_a^{I} \, {\rm d} ~=~& i \, (\gamma^5\gamma^\mu){}_a {}^b \, 
\,  \partial_\mu \psi_b^{I}  ~~~. \cr
} \label{eq:VIspecific}
\ee
where
\be
\sigma^{\mu\nu} = \frac{i}{2}[\gamma^{\mu},\gamma^{\nu}],~~~F_{\mu\nu} = \partial_{\mu}A_{\nu} - \partial_{\nu}A_{\mu}.
\ee
and our conventions for the gamma matrices are as in Appendix A of~\cite{Gates:2009me}.

\subsection{Algebra}\label{CCCVReductionNequals4Algebra}
Using the shorthand
\be\eqalign{
\chi = ( A^I, B^I, F^I, G^I, {\rm d}, \psi^J_c, \lambda_c) ,
}\ee

\noindent the algebra can be written
\be\label{eq:00termsspecific1} \eqalign{
 \{{\rm D}_a, {\rm D}_b\} \chi &=   2 i (\gamma^{\mu})_{ab}\partial_\mu \chi,~~~\{{\rm D}_a, {\rm D}_b\}A_\nu = 2 i (\gamma^{\mu})_{ab}F_{\mu\nu}
}\ee

\noindent and
\be\label{eq:IJtermsspecific1} \eqalign{
\{{\rm D}_a^I, {\rm D}_b^J\} A^K = & 2 i \delta^{IJ} ( \gamma^\mu)_{ab}  \partial_\mu A^K - 2 \epsilon^{IJK} (\gamma^5)_{ab} {\rm d} + \cr
&-2 Z^{IJKM}[i C_{ab} F^M + (\gamma^5)_{ab}G^M],  \cr
\{{\rm D}_a^I, {\rm D}_b^J\} B^K = & 2 i \delta^{IJ} ( \gamma^\mu)_{ab}  \partial_\mu B^K + 2~i \epsilon^{IJK} C_{ab} {\rm d},  \cr
\{{\rm D}_a^I, {\rm D}_b^J\} F^K = & 2 i \delta^{IJ} ( \gamma^\mu)_{ab}  \partial_\mu F^K + 2 \epsilon^{IJK} (\gamma^5\gamma^\mu)_{ab} \partial_\mu {\rm d} + \cr
&+ 2 Z^{IJKM}[ -i C_{ab} \square A^M + (\gamma^5\gamma^\mu)_{ab} \partial_\mu G^M]  \cr
\{{\rm D}_a^I, {\rm D}_b^J\} G^K = & 2 i \delta^{IJ} ( \gamma^\mu)_{ab}  \partial_\mu G^K - 2 \epsilon^{IJK} (\gamma^5\gamma^\mu)_{ab} \partial^\nu F_{\mu\nu} +\cr
 &- 2 Z^{IJKM}[(\gamma^5)_{ab} \square A^M  + (\gamma^5\gamma^\mu)_{ab} \partial_\mu F^M]\cr
}\ee

\be\label{eq:IJtermsspecific2} \eqalign{
\{{\rm D}_a^I, {\rm D}_b^J\} {\rm d} =& 2 i \delta^{IJ} ( \gamma^\mu)_{ab}  \partial_\mu {\rm d} + \cr
  &+ 2 \epsilon^{IJK} ((\gamma^5)_{ab} \square A^K - i  C_{ab} \square B^K + (\gamma^5\gamma^\mu)_{ab} \partial_\mu F^K) \cr
\{{\rm D}_a^I, {\rm D}_b^J\} A_\nu = &  2 i \delta^{IJ} ( \gamma^\mu)_{ab}  F_{\mu\nu} + \cr
  &+ 2 \epsilon^{IJK} (i C_{ab} \partial_\nu A^K + (\gamma^5)_{ab} \partial_\nu B^K - (\gamma^5\gamma_\nu)_{ab} G^K) \cr
\{{\rm D}_a^I, {\rm D}_b^J\} \lambda_c = &  2 i \delta^{IJ} ( \gamma^\mu)_{ab}  \partial_\mu \lambda_c + i \epsilon^{IJK} [-C_{ab} (\gamma^\mu)_c^{~d} + (\gamma^5)_{ab}(\gamma^5\gamma^\mu?)_c^d? + \cr &~~~~~~~~~~~~~~~~~~~~~~~~~~~~~~~~~+(\gamma^5\gamma^\nu?)_ab?(\gamma^5\gamma_\nu\gamma^\mu?)_c^d?] \partial_\mu \psi_d^K \cr
\{{\rm D}_a^I, {\rm D}_b^J\} \psi_c^K = &  2 i \delta^{IJ} ( \gamma^\mu)_{ab}  \partial_\mu \psi_c^K - i \epsilon^{IJK} [-C_{ab} (\gamma^\mu)_c^{~d} + (\gamma^5)_{ab}(\gamma^5\gamma^\mu?)_c^d? + \cr &~~~~~~~~~~~~~~~~~~~~~~~~~~~~~~~~~~~~~+(\gamma^5\gamma^\nu?)_ab?(\gamma^5\gamma_\nu\gamma^\mu?)_c^d?]\partial_\mu \lambda_d  + \cr  &~~~~~~~~~~~~~~~~~~- i Z^{IJKM}[C_{ab}(\gamma^\mu?)_c^d? + (\gamma^5)_{ab}(\gamma^5\gamma^\mu?)_c^d? +  \cr &~~~~~~~~~~~~~~~~~~~~~~~~~+(\gamma^5\gamma^\nu?)_{ab}?(\gamma^5\gamma_\nu\gamma^\mu?)_c^d?] \partial_\mu \psi_d^M  
}\ee 

\noindent and for the cross terms

\be\label{eq:crosstermsspecific1}\eqalign{
\{{\rm D}_a, {\rm D}_b^I\} A^J = & 2 i \epsilon^{IJK}C_{ab} F^K  \cr
\{{\rm D}_a, {\rm D}_b^I\} B^J = & 2 i \epsilon^{IJK}C_{ab} G^K \cr
\{{\rm D}_a, {\rm D}_b^I\} F^J = & 2 i \epsilon^{IJK} C_{ab} \square A^K \cr
\{{\rm D}_a, {\rm D}_b^I\} G^J = & 2 i \epsilon^{IJK}C_{ab} \square B^K \cr
\{{\rm D}_a, {\rm D}_b^I\}\lambda_c = &0
}\ee
\be\label{eq:crosstermsspecific2}\eqalign{
\{{\rm D}_a, {\rm D}_b^I\}{\rm d} = &0 \cr
\{{\rm D}_a, {\rm D}_b^I\}A_\nu = & 2 i C_{ab} \partial_\nu A^I - 2 (\gamma^5)_{ab} \partial_\nu B^I \cr
\{{\rm D}_a, {\rm D}_b^I\} \psi_c^J = & 2i \epsilon^{IJK}C_{ab}(\gamma^\mu?)_c^d? \partial_\mu \psi_d^K 
}\ee
where
\be
   Z^{IJKM} \equiv \delta^{IM}\delta^{JK} - \delta^{IK}\delta^{JM}
\ee

\end{document}